\documentclass[useAMS,usenatbib,fleqn]{mnras}
\usepackage{soul}

\usepackage{bm}
\usepackage[T1]{fontenc}
\usepackage[utf8]{inputenc}
\usepackage{multicol}
\usepackage{times,amsmath,amssymb,longtable,breqn}
\usepackage[varg]{txfonts}
\usepackage{grffile} 
\usepackage{graphicx}
\usepackage{color}
\usepackage[dvipsnames]{xcolor}
\usepackage{hyperref}
\usepackage{comment}

\usepackage{enumitem}
\setlist[itemize]{leftmargin=6mm}
\setlist[enumerate]{leftmargin=6mm}

\usepackage{ulem}
\newcommand\U[1]{{\,\rm #1}}

\newcommand\E[1]{\times10^{#1}}

\newcommand\al{\alpha}
\newcommand\dl{\delta}
\newcommand\om{\omega}

\newcommand\rs[1]{_\mathrm{#1}}
\newcommand\Esn{E\rs{sn}}
\newcommand\rhoism{\rho\rs{0}}
\newcommand\Mej{M\rs{ej}}
\newcommand\rhoej{\rho\rs{ej}}
\newcommand\Rshell{R\rs{shell}}
\newcommand\Vshell{V\rs{shell}}
\newcommand\Mshell{M\rs{shell}}
\newcommand\vt{v\rs{t}}
\newcommand\Rsnr{R\rs{snr}}
\newcommand\Rpwn{R\rs{pwn}}

\newcommand\Qpwn{Q\rs{pwn}}
\newcommand\Rch{R\rs{ch}}
\newcommand\tch{t\rs{ch}}
\newcommand\Lch{L\rs{ch}}
\newcommand\Lz{L\rs{0}}
\newcommand\tauz{\tau\rs{0}}
\newcommand\lgLz{\log_{10}(\Lz/\Lch)}
\newcommand\lgtauz{\log_{10}(\tauz/\tch)}
\newcommand\lgtauzD{\log_{10}\left(\frac{\tauz}{\tch}\right)}
\newcommand\Lztauz{\Lz\tauz}
\newcommand\lgLztauz{\log_{10}(\Lztauz/\Esn)}
\newcommand\lgLztauzD{\log_{10}\left(\frac{\Lztauz}{\Esn}\right)}
\newcommand\LztauzEFF{(\Lztauz)\rs{EFF}}
\newcommand\lgLztauzEFF{\log_{10}[\LztauzEFF/\Esn]}

\newcommand\lgEeff{\lambda\rs{E}}
\newcommand\lgCF{\log_{10}\hbox{CF}}
\newcommand\lgCFlow{\log_{10}\hbox{CF}\rs{low}}
\newcommand\Vpwnscal{\mathcal{V}_0}
\newcommand\Pouter{P\rs{outer}}
\newcommand\aouter{a\rs{outer}}
\newcommand\Psedov{P\rs{Sedov}}
\newcommand\tini{t\rs{ini}}
\newcommand\tbegrev{t\rs{beg,rev}}
\newcommand\tfstsh{t\rs{1st,sh}}
\newcommand\tsndsh{t\rs{2nd,sh}}
\newcommand\Rbegrev{R\rs{beg,rev}}
\newcommand\Vbegrev{V\rs{beg,rev}}
\newcommand\Mbegrev{M\rs{beg,rev}}
\newcommand\Qbegrev{Q\rs{beg,rev}}
\newcommand\Minf{M_{\infty}}
\newcommand\Qinf{Q_{\infty}}
\newcommand\Rmax{R\rs{max}}
\newcommand\Rmin{R\rs{min}}
\newcommand\timplo{t\rs{implo}}
\newcommand\Vimplo{V\rs{implo}}

\newcommand\Rapprox{R\rs{appr}}
\newcommand\CRz{C_{R,0}}
\newcommand\CRinf{C_{R,\infty}}
\newcommand\Qapprox{Q\rs{appr}}
\newcommand\CQz{C_{Q,0}}
\newcommand\CQinf{C_{Q,\infty}}

\begin{document}
\label{firstpage}
\pagerange{\pageref{firstpage}--\pageref{lastpage}}

\title[Reverberation of PWNe (II)]{Reverberation of pulsar wind nebulae (II):\\
Anatomy of the ``thin-shell'' evolution}
\author[Bandiera et al.]
{R. Bandiera$^{1}$, N. Bucciantini$^{1,2,3}$, J. Mart\'in$^{1,4,5}$, 
B. Olmi$^{1,7}$, D. F. Torres$^{4,5,6}$ \thanks{All authors have contributed equally to this work.}  \\
$^{1}$ INAF - Osservatorio Astrofisico di Arcetri, Largo E. Fermi 5, I-50125 Firenze, Italy \\
$^{2}$ Dipartamento di Fisica e Astronomia, Universit\`a degli Studi di Firenze, Via G. Sansone 1, I-50019 Sesto F. no (Firenze), Italy \\
$^{3}$ INFN - Sezione di Firenze, Via G. Sansone 1, I-50019 Sesto F. no (Firenze), Italy \\
$^{4}$ Institute of Space Sciences (ICE, CSIC), Campus UAB, Carrer de Can Magrans s/n, 08193 Barcelona, Spain \\
$^{5}$ Institut d'Estudis Espacials de Catalunya (IEEC), Gran Capit\`a 2-4, 08034 Barcelona, Spain \\
$^{6}$ Instituci\'o Catalana de Recerca i Estudis Avan\c cats (ICREA), 08010 Barcelona, Spain \\
$^{7}$ INAF - Osservatorio Astronomico di Palermo, Piazza del Parlamento 1, I-90134 Palermo, Italy
}

\date{Accepted 2022 December 26. Received 2022 November 4; in original form 2022 July 22}
\maketitle

\pubyear{2023}

\begin{abstract}
During its early evolution, a pulsar wind nebula (PWN)  sweeps the inner part of the supernova ejecta and forms a thin massive shell.
Later on, when the shell has been reached by the reverse shock of the supernova remnant, the evolution becomes more complex, in most cases reverting the expansion into a compression: this later phase is called ``reverberation''.
Computations done so far to understand this phase have been  mostly performed in the thin-shell approximation, where the evolution of the PWN  radius is assimilated to that of the swept-up shell under the effect of both the inner pressure from the PWN, and the outer pressure from the supernova remnant.
Despite the thin-shell approach seems rather justifiable, its implementations have so far been inaccurate, and its correctness, never tested. 
The outer pressure  was naively assumed to be scaled according to the Sedov solution (or a constant fraction of it) along the entire evolution. 
The thin-shell assumption itself fails along the process, being the shell no longer thin in comparison with the size of the PWN.
Here, through a combination of numerical models, dimensional arguments, and analytic approximations, we present a detailed analysis of the interaction of the PWN with the supernova remnant. 
We provide a new analytic approximation of the outer pressure, 
beyond the Sedov solution, and a revised ``thin-shell'' able to reproduce results from numerical simulations. Finally, we compute the efficiency by which the  PWN is compressed during reverberation over a wide population of sources.
\end{abstract}

\begin{keywords}
radiation mechanisms: non-thermal -- pulsar: general -- method: numerical -- ISM: supernova remnants  
\end{keywords}

\section{Introduction}
\label{sec:intro}


Pulsar Wind Nebulae (PWNe) are among the most important high-energy
astrophysical sources in the Universe \citep{Gaensler_Slane06a,Hester:2008}.
Their close proximity makes them a  unique laboratory where to investigate high-energy processes: from particle acceleration \citep{Arons:2012, Sironi2017}, to non-thermal emission \citep{Westfold59a,Weisskopf1978,Kennel_Coroniti84b}, from the fluid-dynamics and physical conditions
of relativistic outflows \citep{Contopoulos_Kazanas+99a, Spitkovsky:2006}, to the properties of their
interaction and confinement by the ambient medium \citep{Rees:1974,Kennel_Coroniti84b}. 
They act as an imager for the pulsar wind, providing us a unique way to gain
knowledge on magnetospheric properties of compact objects that will otherwise be
unobservable \citep{Gaensler_Slane06a}. 
PWNe might be one of the major contributor to leptonic antimatter in the Galaxy \citep{Blasi:2011, Amato:2018}. 
The techniques, methods and tools developed for their investigation have in the past
proved useful in many other contexts from active galactic nuclei black holes \citep{Komissarov:2001}
to gamma-ray bursts \citep{Bucciantini_Thompson+07a}. 
The Crab nebula is one of the most and best studied object of the entire sky \citep{Hester:2008}. 
Moreover they constitute one of the main targets of TeV gamma-ray observatories like LHAASO \citep{Aharonian2020} and the Cherenkov Telescope Array (CTA), and are likely to dominate the background of galactic diffuse TeV emission \citep{de-Ona-Wilhelmi:2013, Klepser:2013, HESS-PWN-2018, Fiori:2021} and generate most of the expected source confusion in surveys
\citep{Mestre2022}.

So far, PWNe have been mostly described with time-dependent one-zone (also named $0+1$) models \citep{Gelfand_Slane+09a,Bucciantini_Arons+11a,Martin_Torres+12a,Torres2013,Torres:2014,Martin2022j}, especially when a large number of systems, a wide parameter space, or an extended evolution is investigated.
One-zone models represent the PWN as a uniform bubble interacting with the surrounding supernova remnant (SNR) and subject to energy (adiabatic and radiative) and particles losses \citep{Reynolds:1984}. 
The PWN radius ($R$) is  associated with that of the massive shell accumulating at the PWN boundary, forming a thin layer of thickness $\Delta R \ll R$. The validity of that condition was originally demonstrated at early times \citep{Jun1998}, but it can be shown it remains valid at later ones, before the PWN starts to interact with the SNR.
All of this will be discussed in detail in Section \ref{sec:pwn}.
The thin-shell approximation is --at least in principle-- well motivated at early times, when the PWN still interacts with unshocked, and freely expanding SNR ejecta. 
Both self-similar solutions \citep{Chevalier1982} and numerical models \citep{van-der-Swaluw:2001, Bucciantini:2003} indeed show that the mass swept up by the PWN form a rather thin-shell of relatively cold material. 
And, apart from multi-D instabilities that could affect the integrity of this shell \citep{Blondin:2001,Porth:2014a, Olmi_Torres:2020}, it will be preserved also at later times \citep{Bucciantini:2004a}.
Observations however suggest that magnetic field might prevent the growth of small scale instabilities and allow the PWN to retain a more coherent shape \citep{Ma_Ng+16a}.

A more realistic representation of the highly dynamical --and complex-- interaction of the PWN with the SNR would require multi-D models.
However, they need huge computational resources, especially in 3D, so that they have only been run for selected, and time limited, studies \citep{Porth:2014, Olmi:2016,Kolb:2017,Olmi_Bucciantini:2019}.
These studies have shown that at such early ages, the pressure and even the magnetic field within the PWNe are relatively 
uniform on large scales, explaining the success of one-zone models in describing the spectral energy distribution we measure from these objects.

Thus, despite their limitations, one-zone models still represent the best tools to describe the evolution of the PWN+SNR system. 
They were proved to be robust in describing the first phase of the PWN+SNR interaction, when the PWN expands inside the SNR with a mild acceleration \citep{van-der-Swaluw:2001, Bucciantini:2003, Gelfand_Slane+09a,Martin_Torres+12a}. Moreover they are the only possible way to evolve the PWN+SNR system coupled to the spectral properties of the PWN particles, while hydrodynamic models cannot account for the evolution of the particle spectra, that requires the Particle In Cell (PIC) approach in more than 1D \citep{Sironi:2009,Sironi2017}, whose necessity in terms of spatial resolution makes it not suitable to approach the present problem.

However, {\it all good things} come to an end. The first phase of the PWN+SNR interaction terminates when the reverse shock (RS hereafter) of the SN explosion reaches the PWN, starting to interact with the swept-up shell.
Due to the combined effect of mass accretion and thermal pressure of the shocked SNR medium, the shell decelerates, in some cases leading to an efficient compression of the PWN. 
As a consequence, the internal pressure of the PWN starts to increase and it eventually becomes high enough to let the system re-expand. 
This phase of contraction and re-expansion is called \textit{reverberation} \citep{Blondin:2001, van-der-Swaluw:2003, Bucciantini:2003}.
The compression may act as an energizer for the emitting particles: the nebular magnetic field is enhanced and particles heated, producing a strong increase of the emission at all the wavelengths, changing significantly the spectral properties of the source \citep{Gelfand_Slane+09a}, and even leading to superefficiency (phases in which the emitted luminosity at given frequency range 
exceeds the spin-down power at the time, \citealt{Torres2018b, Torres2019}). The critical issue is that in reverberation, the rotational power of the pulsar is no longer the energy reservoir of the PWN system, since during a compression it is receiving energy from the environmental interaction.
Much of this depends on the strength of the compression and the relative values of the pressures in the PWN and in the SNR.
The strength of the compression can be quantified with the so called \textit{compression factor} (CF hereafter), defined as the ratio of the maximum (close to the beginning of reverberation) to the minimum radius of the PWN (at its maximum contraction, during the first compression phase), namely:
\begin{equation}
    \mathrm{CF} = \frac{\mathrm{max}[\Rpwn(t)]}{\mathrm{min}[\Rpwn(t)]}\,.
\label{eq:cf}    
\end{equation}
While the pressure inside the PWN is easy to compute, the one in the SNR depends on the dynamics of the interaction itself, and several {\sl ad hoc} prescriptions have been put forward, in general adopting a scaling based on the Sedov solution, e.g., \cite{Gelfand_Slane+09a, Bucciantini_Arons+11a, Vorster:2013, Martin:2016, Torres2019}. 
A preliminary attempt to treat reverberation beyond Sedov approximation  has been recently made by \citet{Fiori:2021}, but with no specific focus in reproducing correctly the dynamical effects of the PWN+SNR interaction.

Despite reverberation may lead to a complete burn-off the pair population, and to significant effects at the morphology level, 
it is not uncommon to see it being simply ignored in the literature. 
The models that do take reverberation effects into account use the thin-shell approximation, in lack of a better treatment.
The main weakness of those thin-shell models  resides precisely in the assumption that the SNR pressure, right outside the shell ($P\rs{outer}$), can be set, at all times, as the central pressure in the Sedov solution, or alternatively, as a constant fraction, scaled according to the Sedov profile, of the pressure downstream the SNR forward shock as derived from \citet{Truelove1999}: two formally different recipes, but giving similar results.
Such an assumption is not justified: it in fact represents only the asymptotic evolution of the pressure, which holds when the SNR is fully relaxed, and formally when the mass of the ejecta becomes negligible with respect to that of the swept-up ambient medium. 
Instead, the most relevant part of the PWN+SNR interaction,
with the onset of the reverberation phase and first compression of the PWN, typically occurs at much earlier times, when the mass of the swept-up material becomes comparable with the mass of the ejecta. 
Furthermore, as we will discuss in the following, numerical models show that the time behavior of the SNR pressure just outside the PWN swept-up shell dramatically deviates from a  Sedov-like trend.

The reverberation phase may dramatically change the spectrum of a source. Thus, a proper evaluation of the many assumptions that go into the thin-shell modeling of the reverberation phase, especially as a function of the parameters describing the PWN+SNR systems, like the energetics of the pulsar or its spin-down timescale, is necessary to understand how well radiative models can deal with reverberation. 
With the advent of the new gamma-ray observatories, building trustworthy models of the late time spectral properties of PWNe becomes more and more pressing. 
The properties and morphology of aged systems might then be very different to what is predicted by a simple naive application of thin-shell models.
The reverberation phase thus deserves an appropriate physical investigation, and in this paper we aim to pave the way to a more accurate parametrization, ultimately, one we will use to improve existing radiative codes.

In \citet[Paper I hereafter]{Bandiera:2020}  we have discussed some of these issues, showing how an  over-simplified description of the pressure profile in the SNR can affect its evolution. 
Here we present both a detailed study of the PWN+SNR evolution, using more sophisticated 1D hydrodynamical codes, and revisit some of the assumption generally used in the literature.
We maintain large part of the standard thin-shell assumptions (pure spherical symmetry, the PWN as an homogeneous bubble, swept-up shell with negligible thickness and internal energy) but we focus instead on an accurate modeling of the outer pressure, at least till the time of the first PWN compression.\\
Before outlining the structure of the paper, let us clarify here one of the main topics that will be discussed, which represents a change of paradigm with respect to most of the treatments of the reverberation presented so far:
\begin{itemize}
    \item The evolution of the PWN+SNR system is strongly dependent on the conditions at the beginning of the reverberation phase, as it was discussed as well in \citet{Bandiera:2020} and \citet{Bandiera:2021}, and for this reason we emphasize that the pre-reverberation phase must be modeled very accurately.
   
    \item The general choice for the pressure outside the PWN swept-up shell of a constant fraction of that derived at the SNR forward shock, and scaled according to the Sedov solution, is not justified because this only holds at much later times.
   
    \item Furthermore, any choice of an outer pressure which is assumed to be independent of the presence of the shell is bound to be wrong, because the interaction of the SNR with the mass shell accumulated around the PWN necessarily affects in turn the outer pressure evolution.

    \item A comparison with models assuming this standard recipe for the outer pressure, in the non-radiative limit, shows that they lead to (even largely) overestimated CFs.
    This is a problem, since overestimating the CF would lead to a different subsequent dynamical evolution history of the PWN.
\end{itemize}
This paper is organized as follows. 
In Sec.~\ref{sec:pplane} we define the parameter plane used in the rest of this work to identify uniquely each PWN+SNR system and its evolution, locating the region populated by most of the systems.
In the Sec.~\ref{sec:pre-revPWN} we describe the evolution of the PWN+SNR system before the onset of reverberation, reviewing and improving analytical models for the initial phase of the interaction.
In Sec.~\ref{sec:numerical} we then describe the numerical scheme used for the simulations, discussing the superiority of lagrangian models to reach the required spatial resolution, as well as the importance of carefully setting the initial conditions. 
Sec.~\ref{sec:shell} is devoted to analyze the interaction of a massive expanding shell with the SNR.
This is a good approximation of the swept-up shell accumulated at the PWN boundary, and responsible for mediating the interaction with the SNR.
We analyze how the SNR behaves in the presence of the shell.
This allows us to describe with much higher accuracy than before the outer pressure acting on the shell.
Then, the full evolution of the PWN+SNR system is described in Sec.~\ref{sec:pwn}. Here we illustrate and discuss the results of numerical models for different energetics and pulsar spin-down time, we critically analyze the thin-shell assumption, and derive approximate formulae for the PWN compression factor.
In Sec.~\ref{sec:rev-thin-shell} we discuss how to apply a thin-shell approach after the onset of reverberation.
Here we present a revised thin-shell model, shaped to agree more closely to the numerical evolution during reverberation.
A comparison of synoptic maps of the compression factor is presented in Sec.~\ref{sec:results}. 
Sec.~\ref{sec:simplified} presents a simplified version of the thin-shell model, able to reproduce the evolution with reasonably good accuracy in the case of small spin-down times.
Our conclusions are finally drawn in Sec.~\ref{sec:end}.
%

\section{The PWN+SNR parameter plane }\label{sec:pplane}

\subsection{Characteristic scales and the $(\tauz/\tch)-(\Lz/\Lch)$ plane}
%
As a first step for the forthcoming discussion, we must define what parameters are fundamental to describe the PWN+SNR evolution and the properties of the reverberation phase.
As already done in Paper I and \citet[Paper 0 hereafter]{Bandiera:2021}, 
we will use characteristic scales for some dimensional quantities, along the lines of those originally introduced by \citet{Truelove1999}, namely:
\begin{eqnarray}
\Rch\!\!\!&=&\!\!\!\Mej^{\,1/3}\rhoism^{\,-1/3},	\\
\tch\!\!\!&=&\!\!\!\Esn^{\,-1/2}\Mej^{\,5/6}\rhoism^{\,-1/3},\\
\Lch\!\!\!&=&\!\!\! \Esn^{\,1/2}\Mej^{\,-5/6}\rhoism^{\,1/3} = \Esn/\tch\,,
\label{eq:chscales}
\end{eqnarray}
where $\rho_0$ is the ambient medium mass density, $\Mej$ the mass of the SNR ejecta and $\Esn$ the supernova energy.
The combined evolution of a PWN+SNR system 
is roughly determined by 
two ratios: $\tauz/\tch$ and $\Lz/\Lch$, where $\Lz$ and $\tauz$ are, respectively, the pulsar initial spin-down luminosity and age.
This holds true not only in the free expansion phase (before the onset of reverberation) but also along the entire PWN evolution: each PWN+SNR system is represented by a point in the 
$(\tauz/\tch)-(\Lz/\Lch)$ plane.
Moving to a discussion scaled on these quantities reduces by 3 ($\Mej, \rho_0, E_{sn}$) the explicit parameters of the problem. 
However, there are other degrees of freedom, among which the braking index, the density distribution of the SNR ejecta or the uniformity of the ISM, that would still make the problem more complex.
%

\begin{figure}
\centering 
	\includegraphics[width=.48\textwidth]{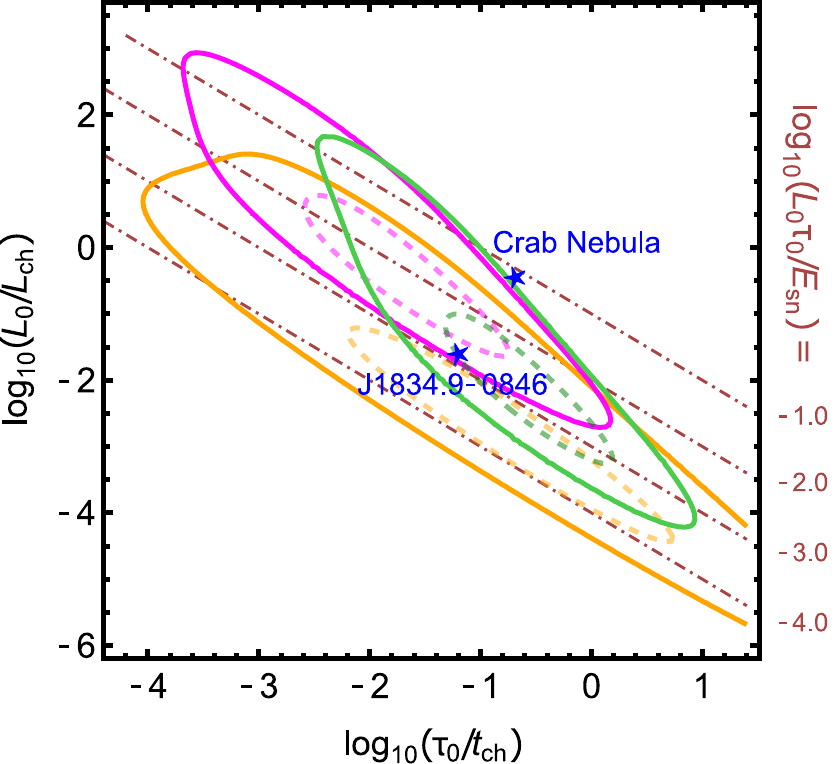}
        \caption{The distributions of pulsars in the $\log_{10}(\tauz/\tch)-\log_{10}(\Lz/\Lch)$ parameter plane, for some synthesized populations.
        For each distribution the dashed curve shows the isolevel enclosing 50\% of the population, while the solid one surrounds the 98\%.
        The population following \citet{FGK:2006} is shown in orange, while that following \citet{Watters:2011} is in magenta.
        For clarity we do not plot here the population from \citet{Johnston:2020}, but it is roughly compatible with that of \citet{Watters:2011}, apart that it is more compact.
        In green color we also show the population that we have devised, which is intermediate between the former two. 
        The two blue stars mark the positions of the Crab nebula and of PSR J1834.9--0846. Finally, the dot-dashed lines show the loci of constant pulsar to SNR energetics.}
\label{fig:moredistributions}	
\end{figure}

\subsection{Properties of the SNe and synthetic pulsar population}

The region of the parameter plane where to concentrate our attention depends on the properties assumed as characteristics of the pulsar plus the SNR populations.
We have in particular considered SNRs to have uniformly distributed ejecta masses in the range $\Mej = [6-20]{\rm M}_\odot$ \citep{Smartt:2009}, and a standard supernova explosion energy of $\Esn=10^{51}\U{erg}$.

Following \citet{Badenes:2010} we have considered a flat power law probability distribution of the ISM density, namely $d\,{\cal P}(\rho_0)/ d \rho_0 \sim {\rho_0}^{-1}$, in the range $\rho_0 = [0.01-10]\,m_p\U{cm^{-3}}$, with $m_p\simeq 1.67\times 10^{-24}\U{g}$ \citep{Berkhuijsen:1987, Magnier:1997, Long:2010, Bandiera_Petruk:2010, Asvarov:2014}.
However, we found that a slight variation of the mass and density ranges does not reflect in an evident modification of the final population.

With the previous prescriptions the characteristic radius $\Rch$ never exceeds $\sim40\U{pc}$, and it is usually much smaller than this.
This justifies our assumption that, during reverberation, the SNR expands in a rather uniform ambient medium.
Larger systems would be so extended that their surface brightness would make them almost undetectable, while their reverberation phase, generally around 1--3\,$\tch$, is delayed up to $\sim 100\U{kyr}$, an age compatible with some of the oldest known runaway PWNe \citep{Kargaltsev2008,Kargaltsev2017}.

For synthesizing a representative pulsar population,
we have assumed that the initial spin-down periods have a Gaussian distribution, characterized by a mean $P_0=100$ ms, a spread $\sigma_{P_0}=80$ ms, and truncated at 10 ms.
This both differs from the distribution by \citet{Watters:2011} for the gamma-ray emitting pulsars (a similar one, less extended, was later found in \citealt{Johnston:2020}) as well as from the well known one for radio pulsars by  \citet{FGK:2006}.
Our choice is somewhat in the middle between the two, covering a large part of the gamma-ray population, plus a significant one of the radio pulsars, limited to the region where an associated PWN can be expected. 
Gamma-ray emitting pulsars seems in fact better representative of the young population of pulsars powering PWNe, conversely to radio pulsars that are more shifted towards old and evolved systems.
Therefore our choice is a good compromise not to emphasize excessively the extremes of both populations.

We take pulsars magnetic fields at birth to follow a log-normal Gaussian distribution, with mean $\log_{10}{(B_0/1 \U{G} )}=12.3$ and $\sigma_{\log_{10}{B_0}} = 0.25$, similar to what was assumed in \citet{FGK:2006} and more recently in \citet{Gullon:2014}.
The braking index has been fixed to $n=2.33$, an intermediate value between the canonical dipole one, $n=3$, and that of PSR 0540-69, $n\simeq 2$ \citep{Manchester:1989}; $n=2.33$ also corresponds to the secular braking index of the Crab pulsar \citep{Lyne:2015, Horvath:2019}.
For a complete discussion on the large range for the few known braking indices please refer to \citet{Parthasarathy:2020}.

In Fig.~\ref{fig:moredistributions} we  show how our population (in green color) locates in the $\log_{10}(\tauz/\tch)-\log_{10}(\Lz/\Lch)$ parameter plane, and how it compares with other possible choices computed using pulsar populations from the literature (\citealt{Watters:2011} in magenta, \citealt{FGK:2006} in orange). 
Please notice that we have in all cases used a braking index of $n=2.33$.
In Fig.~\ref{fig:moredistributions} we also give the position of two reference objects (with blue stars), already introduced and discussed in Paper I: the Crab nebula and J1834.9--0846.
These are a sort of extreme cases from the observational point of view, being the one fed by a very powerful pulsar -- the Crab -- and the other, with spin-down
power smaller by a factor of $\sim50$, representative of a low energetic PWN.
The two are roughly located on the outskirts of our population of interest.
Dot-dashed lines represent constant ratios of PWN+SNR energetics $\Lztauz/\Esn$.
We caution the reader that many of the parameters defining the PWN+SNR population are poorly constrained, 
however we believe that our choice allows us to define a realistic population, compatible with available observations.

\section{PWN pre-reverberation evolution}
\label{sec:pre-revPWN}

This section describes the evolution of the PWN before the onset of the reverberation phase, namely as long as it interacts with the unshocked ejecta. 
Apart from possible cases of exotic pulsars, with an energy comparable to or even larger than the supernova energy itself, the PWN never reaches the outer ejecta steep envelope, and this fact simplifies the modeling.
Moreover in Paper 0 we have investigated the effect of the variation of the parameters shaping the ejecta density profile: the core, with power-law index $\dl$, and the envelope with index $\om$ (see also Eq.~\ref{eq:gendensityprofile}). 
From the observational point of view, these two parameters are hardly constrained, while they are in general considered to vary in the ranges $0\lesssim \dl \lesssim 1$ and $9\lesssim \om \lesssim 12$ \citep{Chevalier:1989,Truelove1999,Matzner:1999,Chevalier2005,Kasen:2010,Bucciantini_Arons+11a,Miceli:2013,Potter2014,Karamehmetoglu:2017,Kurfurst:2020,Meyer:2020,Meyer2021}.
We found that no relevant differences in the evolution of the SNR characteristic curves (reverse shock RS, contact discontinuity CD and forward shock FS) appear for the different profiles of the ejecta core with $\dl\lesssim 0.1$, while the envelope profile for all $\om\gtrsim 9$ is very well described by the asymptotic limit $\om=\infty$. 
Here we then specialize our analysis to the case of a density distribution in the SNR ejecta with a flat core 
and an infinitely steep envelope: $\rhoej(t)=A/t^3$, corresponding to the $\dl=0$ and $\om=\infty$ case of the more general formulae presented in Appendix \ref{sec:Vearly-pwn-evoGEN}.
In the chosen case the parameter $A$ is given by: $A=5\Esn/(2\upi) \,[3\Mej/(10\Esn)]^{5/2}$.

During its early evolution ($t\ll\tauz$) the PWN collects material from the ejecta into a shell with thickness much smaller than the PWN radius \citep{Jun1998}.
As already mentioned, this result is at the basis of the thin-shell approximation, widely used in the literature to describe the evolution both before and during reverberation
\citep[e.g.][]{Reynolds:1984, Bucciantini:2004a, Gelfand_Slane+09a, Martin:2016, Torres:2017}.
The equations for the PWN evolution in the adiabatic regime and in thin-shell approximation are:\footnote{Note 
that in \citet{Gelfand_Slane+09a}, as well as in many 
subsequent papers referring to that work, the equation for the mass conservation has been incorrectly written in a slightly different way, and this has led to underestimate the PWN size typically $\sim20\%$. 
In Eq.~29 of \citet{Gelfand_Slane+09a} (namely, $dM/dt = 4\upi R^2 \rhoej dR/dt$)  the mass increment is evaluated as the product of $\rhoej$ times the volume increment, which would be correct only if the ejecta were at rest.
The correct formula reads $dM/dt =4\upi R^2 \rhoej\left[dR/dt-R/t\right]$.
The problem is solved here by directly using the integral equation, Eq.~\ref{eq:Mswept}. } 
\begin{eqnarray}
\label{eq:magfluxcons}
\frac{d}{dt}\left(4\upi\,P(t)R(t)^4\right)&=&L(t)\,R(t)\,,\\
\label{eq:momentumcons}
\frac{d}{dt}\left(M(t)\frac{dR(t)}{dt}\right)&=&4\upi\,P(t)R(t)^2 +\frac{dM(t)}{dt}\frac{R(t)}{t} \,.
\end{eqnarray}
We indicate with $R(t)$ the radius of the shell (by definition, in this approximation equivalent to that of the PWN), with $M(t)$ the shell mass, $P(t)$ the total pressure at the inner shell boundary and $L(t)$ the pulsar spin-down luminosity at time $t$.
%
By introducing the quantity:
\begin{equation}
Q(t)=4\upi\,P(t)R(t)^4,
\end{equation}
the two equations above can be simplified. One can also notice that $Q(t)$ tends to an asymptotic value for $t\gg\tau_0$. 

\noindent For the flat density profile, the swept-up mass turns out to be:
\begin{equation}
\label{eq:Mswept}
M(t)=\frac{4\upi\,R(t)^3}{3}\frac{A}{t^3}\,.
\end{equation}
%
The time evolution of the pulsar spin-down luminosity from the initial values $(\tau_0,\, L_0)$ is:
\begin{equation}\label{eq:Edot}
L(t) =\Lz \left( 1 + \frac{t}{\tauz} \right)^{-\al},
\end{equation}   
where we have introduced the fading index $\al=(n+1)/(n-1)$. As discussed in the previous section, here we fix $n=2.33$, or equivalently $\al=2.5$.

In Appendix~\ref{sec:Vearly-pwn-evoGEN} we have derived, for a generic bipower-law density distribution of the unshocked ejecta, the 1$^{\mathrm{st}}$-order analytic approximation of the early-time ($t\ll \tauz$) solution of Eqs.~\ref{eq:magfluxcons}--\ref{eq:momentumcons}.
In the case of homogeneous ejecta that solution simplifies to:
\begin{eqnarray}
R_{(1)}(t)&=&\left(\frac{125}{33}\frac{\Lz t^{6}}{4\upi\,A}\right)^{1/5}\left(1-\frac{11}{245}\,\al\frac{t}{\tauz}\right)\nonumber\\
&\simeq&0.7868\,\,\left(\frac{\Lz t^{6}}{A}\right)^{1/5}\,\,\left(1-\frac{11}{245}\,\al\frac{t}{\tauz}\right)\,, \label{eq:solrt_A}\\
Q_{(1)}(t)&\simeq&0.3576\,\,\left(\frac{\Lz^6 t^{11}}{A}\right)^{1/5}\left(1-\frac{176}{245}\,\al\frac{t}{\tauz}\right)\,, \\
P_{(1)}(t)&\simeq&0.07428\left( \frac{\Lz^2 A^3}{t^{13}}\right)^{1/5}\left(1-\frac{132}{245}\al\frac{t}{\tauz}\right)\,.
\end{eqnarray}
Generally only the 0$^{\mathrm{th}}$-order approximations are given, but we found that the 1$^{\mathrm{st}}$-order expansion in $t/\tauz$ leads to 
a higher accuracy in the description of the early time evolution, and should then be preferred to correctly set the initial conditions for numerical modeling.
Incidentally, note that while the 0$^{\mathrm{st}}$-order terms are independent of $\al$, the first-order terms are all linearly dependent on it.

At later times, instead, it is not possible to find an analytic solution and Eqs.~\ref{eq:magfluxcons}-\ref{eq:momentumcons} must be solved numerically.
The asymptotic formulae for later evolution ($t\gg\tauz$ but still before reverberation) are:
\begin{eqnarray}
\label{eq:asimp1}
R_{(\infty)}(t)&\simeq&0.9522\,\Vpwnscal t\,,\\ 
\label{eq:asimp2}
Q_{(\infty)}(t)&\simeq&1.1293\,\Lz\Vpwnscal\tauz^2\,,\\
\label{eq:asimp3}
P_{(\infty)}(t)&\simeq&0.10932\,\frac{A\Vpwnscal^2\tauz}{t^4}\,,
\end{eqnarray}
where $\Vpwnscal=(\Lztauz/A)^{1/5}$.
The quantity $\Vpwnscal\tauz$ can be then expressed in terms of characteristic scales as:
\begin{equation}
    \Vpwnscal\tauz\simeq 1.9111\left(\frac{\Lztauz}{\Esn}\right)^{1/5}\left(\frac{\tauz}{\tch}\right)\Rch\,.
\end{equation}
From the previous equations one may also estimate the value asymptotically approached by the swept-up mass:
\begin{equation}
\label{eq:Masympt}
\Minf\simeq0.9902\left(\frac{\Lztauz}{\Esn}\right)^{\,3/5} \Mej\,.
\end{equation}
We have derived a handy approximation for $R(t)$, which allows one to closely match both the early and the late asymptotic behaviors, as well as to reasonably describe the transition region:
\begin{equation}
\Rapprox(t)= \Vpwnscal\tauz
  \left[ 1+\left(\CRz^{5/6}\frac{t}{\tauz}\right)^{-a}\right]^{-6/(5a)}
  \left[ 1+\left(\CRinf\frac{t}{\tauz}\right)^b \right]^{1/b}\!\!.
  \label{eq:Rapprox}
\end{equation}
The values of the four parameters depend on the value of the fading index.
In the case $\al=2.5$ the best-fit values are: $\CRz=0.7868$, $\CRinf=0.9522$, $a=0.7297$, $b=0.7194$, ensuring at all times an accuracy better than 0.1\%. For other braking indices see Appendix \ref{sec:pwn-evo-long}.
From this, with an accuracy always better than 0.8\%, the expansion parameter reads:
\begin{equation}
m=\left.\frac{V(t)\,t}{R(t)}\right|\rs{appr}=\frac{6/5}{1+\Big(\CRz^{5 /6}t/\tauz\Big)^a}+\frac{1}{1+\Big(\CRinf\,t/\tauz\Big)^{-b}}\,.
  \label{eq:ExpFactapprox}
\end{equation}
In an analogous way we have approximated the quantity $Q(t)$, with an accuracy always better than 0.8\%, as:
\begin{equation}
\Qapprox(t)=\Lztauz\Vpwnscal\tauz
  \left[ \left(\CQz^{5/11}\frac{t}{\tauz}\right)^{-a}+b\left(\frac{t}{\tauz}\right)^{-c}+\CQinf^{-5a/11}\right]^{-11/5a}\!\!,
  \label{eq:Qapprox}
\end{equation}
where $\CQz=0.3576$,  $\CQinf=1.1293$, $a=1.4689$, $b=1.6969$ and $c=0.5820$.
Note that the asymptotic value for $Q(t)$ is finite and holds:
\begin{equation}
\Qinf=\CQinf\Lztauz\Vpwnscal\tauz.
\label{eq:Qasympt}
\end{equation}
Compared with the analytic approximation in \citet{Bucciantini:2004a}, the present one is far more accurate, especially at late times; on the other hand, it is limited to the case $\al=2.5$.
A comparison of the above approximation for $R(t)$ with the fully numerical solutions (presented in Sec.~\ref{sec:numerical})
allows one to test the convergence of the simulations magenta and 
to estimate the error introduced by assuming a thin-shell (see Sec.~\ref{sec:rev-thin-shell}).

The time at which the reverberation begins, hereafter $\tbegrev$, can be identified as the time at which  the PWN and the  RS first touch.
This time can be derived very precisely in the case of a thin-shell approximation for the swept-up shell, while in the real case of a finite thickness the RS takes a small but finite time to propagate through the shell.
For the RS evolution we  use the general formula given by Paper 0, specialized to the present case of $\dl=0$, $\om=\infty$, that is extremely accurate if compared with the numerical models (better than $0.005\,\Rch$) at all values $t<2\,\tch$, but shows a slightly worst agreement (up to $0.03\,\Rch$) close to the RS implosion. 
Since in the present case it will be used to derive other quantities, we have preferred to refine the parameters for a perfect consistence with our numerical models, giving:
%
\begin{equation}\label{eq:RRS}
    R\rs{RS}(t)=\frac{12.49\,(2.411-t/\tch)^{0.6708}(t/\tch)^{1.663}}{1+17.47\,t/\tch+4.918\,(t/\tch)^2}\Rch\,,
\end{equation}
that allows an accuracy in comparison to the lagrangian model always better than $0.002\,\Rch$ up to $t=2.411\,\tch$, the RS implosion time in the absence of a PWN for the chosen configuration of the ejecta. 
The quantity $\tbegrev$ is in general a function of both $\Lz/\Lch$ and $\tauz/\tch$ and its evaluation requires a numerical derivation of the intersection between the curves for the PWN and RS evolution.
Anyway, one should notice that for $\tauz/\tch\ll 1$, it is a function of only $\Lztauz/\Esn$ while, for  $\tauz/\tch\gg 1$, it is a function of only $\Lz/\Lch$.
It is also easy to show that several quantities at $\tbegrev$ can be expressed as functions of $\tbegrev/\tch$ only.
This is for instance the case of the RS radius (and of the PWN as well), of the amount of swept-up mass, of the velocity of the RS as well as  of  the matter downstream to the RS, and of its thermal pressure.

\section{Numerical Scheme for PWN+SNR evolution}
\label{sec:numerical}

In order to follow the evolution of a PWN+SNR system from a few years after the SN explosion to the typical time of the reverberation,
around  1–2\,$\tch$, meaning  $10^3-10^4$ yrs for most of the systems, we must allow a large dynamical
range in terms of radii.
On top of this, the injection of a relativistic wind from the pulsar, and the related requirement to resolve the dynamics of the wind termination shock and of the PWN, in the correct relativistic regime, makes such simulations even more demanding.

Using the PLUTO code \citep{Mignone2007}, we have verified that a spatial resolution good enough to resolve the structure of the swept-up shell, as well as the PWN CD with the shocked ejecta, is required to correctly simulate, even in 1D, the dynamics of  reverberation and to reliably compute the maximum compression of the PWN (i.e. the CF).
Unresolved shells lead to underestimate the CF even by a factor 3, especially for those systems where the compression is strong. 
The issue is far less critical for systems harboring an energetic pulsar, where the  compression in the reverberation phase is mild (CF$\,\sim2$--$5$).
 
Lagrangian schemes, following the evolution of fluid particles, are instead optimal for this kind of problem, where the 1D geometry prevents dynamical mixing and preserves the identity of mass shells in time. 
Moreover, in a lagrangian scheme it is possible to model the PWN as an inner boundary condition and to evolve only the SNR, in the non-relativistic regime. 
This allows one to efficiently sample the parameter space of the PWN+SNR interaction in a reasonable timescale.
In Paper 0 we have presented a lagrangian 1D hydrodynamic code for the evolution of a SNR in the uniform ambient medium, considering different models for the ejecta structure.
For the present problem we have used the same code, modifying it to include the PWN inside the remnant.
The details of our lagrangian scheme, as well as the comparison with analytical solutions for the SNR evolution, can be found in Paper 0 and are not repeated here.

In the following, the inner PWN is modeled using two different implementations of our code.
In the first case we consider a massive thin-shell in ballistic expansion from the center and from the SN explosion time. Ballistic expansion ends when the shell is reached by the RS, and then continues its evolution acted upon by the pressure of the SNR only; this has been done by setting $\Delta m_{1/2}$, defined in Paper 0, equal to the mass of the thin-shell.
As we will show in the following, this approach is mostly suited for PWNe with a small $\tauz$, and allows for a first analysis of the dynamical reaction of the SNR to the presence of a central PWN, a very important ingredient to build up semi-analytic thin-shell models that are more physically grounded.
In the second case, the PWN is modeled as a piston of variable pressure acting on the first lagrangian shell.

The pressure, as long as one neglects the energy losses due to synchrotron emission, inverse Compton scattering (ICS) radiation, or particle diffusion, is ruled by Eq.~\ref{eq:magfluxcons}, and is explicitly computed as:
\begin{equation}
\label{eq:Ppwn}
P(t) = \frac{1}{4\upi R^{4}(t)}\int_0^{t} L(t') R(t'){\rm d}t'\,,
\end{equation}
where the PWN radius $R(t)$ is set equal to the radius of the first interface $r_{1/2}$.

We verified that the modifications we introduced in the lagrangian model, with respect to its original structure of Paper 0,  do not affect the results of the SNR evolution in the high $\om$ case; we found a perfect coincidence with results presented in Paper 0, for the whole SNR region not yet causally affected by the interaction with the PWN.

An important detail of the models is the setting of the initial conditions.
As for the initial time, we have found numerically good results already by setting $\tini=2\times 10^{-3}\tch$.
In the ballistic shell case, the initial radius and velocity of the shell have been simply obtained by imposing the shell mass to be equal to the swept-up mass of the ejecta, and that the expansion time of the shell is equal to that of the SNR.
Initialization in the piston case is more complex. The initial radius of the piston has been evaluated using the early-time analytic expansion for the PWN, while the initial pressure comes from the analytic approximation of Eq.~\ref{eq:Ppwn}.
In the piston model we then do not introduce any thin-shell, since a shell with a small but finite width is self-generating.
However, in this way a very small mass deficiency is introduced, because before $\tini$ the piston has not collected any mass.
The effect of this choice is usually minor, but for a more detailed analysis of the effects of the initial conditions we refer to Sec.~\ref{sec:rev-thin-shell}, in which numerical data and semi-analytic thin-shell models are compared.

\section{A massive shell interacting with the SNR}
\label{sec:shell}

Before tackling the full problem of the interaction of a PWN with the SNR shocked material, let us discuss here a relatively simpler one: the interaction of a thin, linearly expanding, massive shell with the SNR.
Let $\Mshell$ be the shell mass, while for simplicity we take its expansion age to be equal to the SNR age.
As we will better see in the next sections, this problem is a close cousin of that of a PWN with a very small spin-down time at birth, in which case the entire pulsar energy is converted in the formation of the expanding mass shell in a small time, while the internal energy of the inner hot bubble eventually vanishes.\\
Already by this simpler approach we can outline some of the main problems intrinsic to previous models of PWN reverberation:
\begin{enumerate}[label={\arabic*.}]
    \item The evolution during reverberation is strongly dependent on the conditions at the beginning of this phase, and due to this the modeling of the earlier phase must be very accurate; therefore the formulae introduced in our Sec.~\ref{sec:pre-revPWN}, even if not too different from previous ones, are relevant to correctly model the further evolution. 
    \item For the above reason, even if the trajectories of the characteristic curves, as computed in Paper 0, appear rather similar to the original \citet{Truelove1999} ones (at least if we limit the comparison to the case considered here with $\dl=0$ and $\om=\infty$), their respective effects on the subsequent evolution differ considerably.
    \item Furthermore, most of the models published so far use, for the outer pressure, a scaling based on the Sedov solution.
    Instead, with the help of numerical models, we will show that the use of a Sedov solution for the outer pressure is not only unjustified, but also gives grossly incorrect values. This because: (i) the Sedov solution cannot be used to model early phases of the SNR evolution, even referring to ejecta material; (ii) the interaction of the SNR with the mass shell accumulated around the PWN further modifies the outer pressure evolution, and it is essential to take also this into account.
\end{enumerate}
%
\subsection{Evolution of the shell, effective energetics}

The dynamical evolution of the SNR, even in the absence of any inner interaction, is quite complex by itself.
At very early phases it approaches a self-similar solution, when the RS still moves through the envelope \citep{Chevalier1982}, provided that the RS remains  close to the CD \citep{Hamilton_Sarazin:1984}. 
The asymptotic self-similar regime, well described by the Sedov solution \citep{Sedov1946} is then only approached at very late times. 
On the contrary, the intermediate phase, starting from the time at which the RS is no longer closer to the CD and moves toward the center of the explosion, the hydrodynamic evolution of the SNR structure becomes very complex, and requires numerical models to be properly described.
For instance, after the RS has reached the center, a reflected shock starts moving outwards, and further reflections on the various layers (like the CD and the FS) will follow.
All this is implicitly assuming spherical symmetry; while in a real 3D case the efficiency with which the reflected shock is formed may be lower, but the evolution will be more structured.

In the shell+SNR models discussed here, $\tbegrev$  corresponds to the time at which the RS reaches the shell.
The subsequent evolution of the shell size will depend on how effective the pressure of the SNR shocked medium will be, at the beginning to slow down its expansion, and then to revert it into a contraction: in fact, the shell will always reverse its motion, provided the original shell energy roughly does not exceed the SN energy itself.
It is clear that for this evolution only the pressure right outside the shell matters.
On the other hand, in order to determine the pressure at the interface, models must consistently describe the evolution of the whole SNR.
Also in this case, the described scenario assumes spherical symmetry.
In a real 3D model, even in the case of spherically symmetric initial conditions, instabilities may appear at the interface to break the symmetry out.
However, these possible effects will not be considered in the present work.

With the set of characteristic units introduced above, and fixing $\om=\infty$, all models for the shell+SNR interaction form a one-parameter family, depending on the ratio $\Mshell/\Mej$, instead of $L_0$ and $\tauz$.
However, in order to establish a closer connection to the PWN+SNR models that we will treat in the following, let us introduce an effective PWN energy, $\LztauzEFF$, such that a PWN with this energy, and an arbitrarily small\footnote{This is because we have already seen that the asymptotic swept-up mass (Eq.~\ref{eq:Masympt}) is approached only at times much larger than $\tauz$ and, in order to be as much general as possible in our description, this requirement is certainly satisfied considering very small $\tauz$.} $\tauz$, will eventually collect the same $\Mshell$ from the unshocked ejecta.
For this we have used the asymptotic formula that relates the final shell mass and pulsar energetics (Eq.~\ref{eq:Masympt}), to obtain: 
\begin{equation}\label{eq:forLambda}
\LztauzEFF/\Esn\simeq 1.0166\,(\Mshell/\Mej)^{5/3}.
\end{equation}
This is a considerable simplification with respect to the complete case of the PWN+SNR interaction, where the family of solutions has two independent parameters, $\Lz$ and $\tauz$.
For all cases with non negligible $\tauz$, one may find that $\LztauzEFF$ is always smaller than the actual $\Lztauz$, even though usually by a small amount (see  Sec.~\ref{sec:rev-thin-shell}).
By defining  the following effective energetic magnitude:
\begin{equation}\label{eq:forLambdaL}
\lgEeff=\lgLztauzEFF, 
\end{equation}
the models that we have computed span a wide range of $\lgEeff$, from $-10$ to $-0.5$.

\begin{figure} 
\centering
	\includegraphics[width=.47\textwidth]{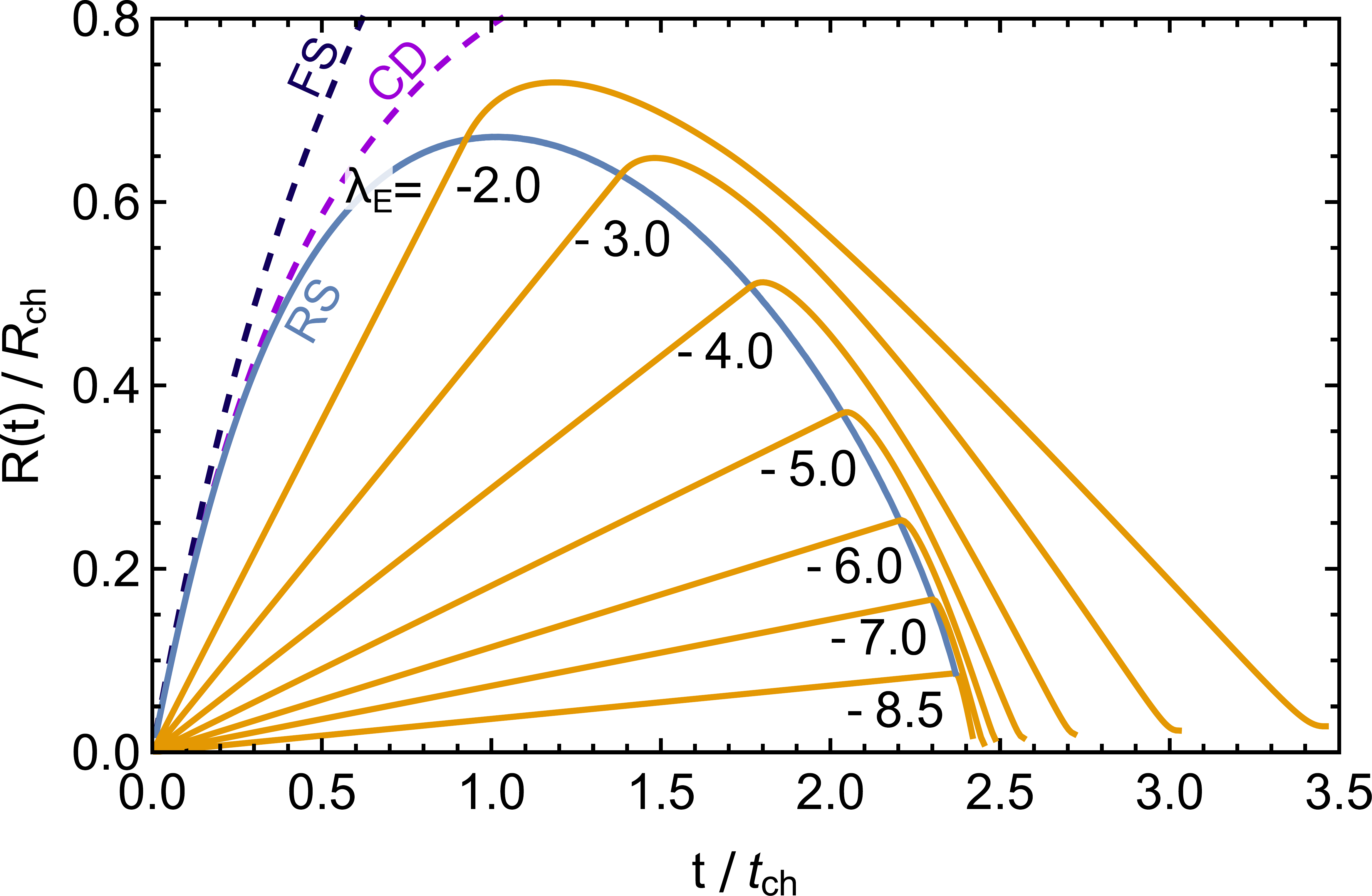}
        \caption{Time evolution of the shell size for models with different $\lgEeff$ (=$\lgLztauzEFF$) values, as obtained from numerical simulations of the interaction between the shell and the SNR. The RS position is indicated by a blue solid line, while the SNR CD and FS by dashed lines, in purple and dark blue colours respectively. The models for the shell are shown as solid orange lines, labelled with their respective values of $\lgEeff$.
        }
\label{fig:diffShells}	
\end{figure}
Fig.~\ref{fig:diffShells} gives an overall scenario  of the evolution of the shell, for a set of representative values of $\lgEeff$: one may notice that it expands linearly, with a velocity dependent on the value of $\lgEeff$, until the shell is reached by the RS; after then the shell expansion is reverted into an implosion, with  efficiency dependent on the energy parameter. 
It is also easy to see that increasing $\lgEeff$ causes the reverberation to begin earlier and end later.
The evolution of CD and FS, represented in the plot by dashed lines, looks almost independent of $\lgEeff$ for lower energies like those present in the plot, but  it also changes considerably in the case of more energetic models.

\subsection{Pressure of the ejecta}
\label{subsec:pressureej}

In several published thin-shell models for the PWN+SNR interaction, and in lack of a better prescription, see e.g., \cite{Gelfand_Slane+09a, Bucciantini_Arons+11a, Vorster:2013, Martin:2016, Torres:2017,Bandiera:2020}, the SNR pressure at the interface with the shell has been imposed to be a constant fraction of that at the FS, as evaluated from the Sedov solution.
This assumption is not well motivated, for various reasons: the range of times at which the shell+SNR interaction takes place is around $\tch$, and extends at most to a few $\tch$, namely it happens well before the Sedov regime has been reached. 
Moreover, the shell interacts with the material in the ejecta, while the profiles provided by the Sedov solution refer to the shocked ambient medium.
In addition, the dynamic thrust of the shell onto the SNR material drives a further reflected shock, so that even a pressure derived from a correct model for a ``pure'' SNR would not be valid here, since the pressure structure in the SNR itself is modified by the interaction with the shell.
For all these reasons we have preferred to compute numerically a large set of shell+SNR models, with the aim of investigating the dependence on $\lgEeff$ of the results, of obtaining an interpolated model from them, and of using this as a better proxy to estimate the dynamical action from the SNR.

Another common assumption in the aforementioned models is that the SN ejecta continue to be swept-up by the shell even beyond the beginning of the reverberation phase, provided that the expansion velocity of the shell is larger than that of the ejecta lying immediately outside i.e., $v_{ej}<v$.
Also this approach is not justified.
The shell of swept-up material formed by a PWN during the pre-reverberation phase is thin for the coexistence of two conditions: (i) the swept-up material (unshocked ejecta) is cold; (ii) the difference between the velocity of the shell and that of the outer material is small.
As a consequence the entropy of the material inside the shell is small and, being the shell in pressure equilibrium with the 
surroundings, this implies that its density is large.
During reverberation, the conditions are completely different; the SNR material hitting the shell is already hot (shocked material), and in addition its velocity relative to that of the shell is high.
Thus, in the absence of very efficient radiative processes, there is no way by which this material can stick to the shell, but it will more likely form a reflected shock.
In the present work all these details are not simply ``assumed'', but they directly result from the numerical models. A more quantitative discussion is found in Sec.~\ref{sec:pwn}.

\begin{figure}
\centering
	\includegraphics[width=.47\textwidth]{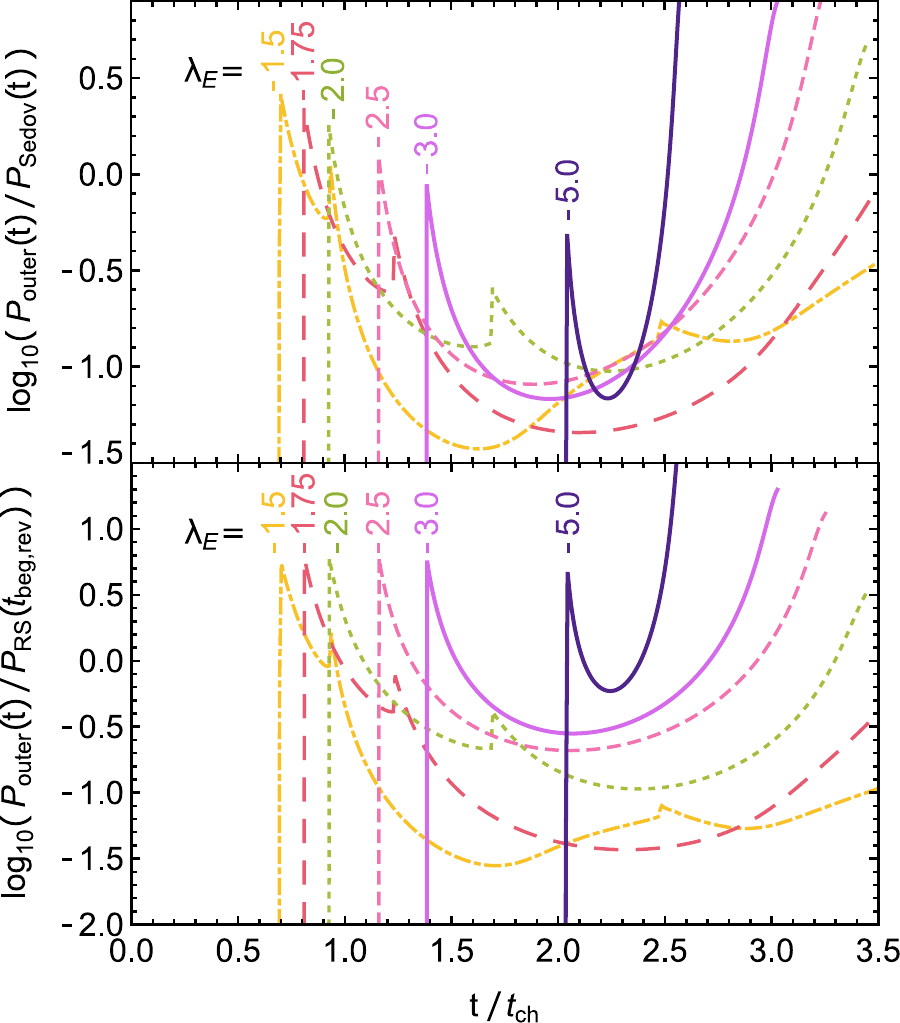}
        \caption{Evolution of the outer pressure, scaled with: $\Psedov$ from Eq.~\ref{eq:psedov} (upper panel); the pressure downstream of the RS at $\tbegrev$ (bottom panel). Pressure profiles are shown in both cases for some representative values of $\lgEeff$, as obtained from numerical solutions.
        Differently from what was assumed in previous works, it is evident here that these quantities are both far from being constant in time. In addition, a secondary bump occurs in the two curves with the highest $\lgEeff$ values: this is the sign of a reflected shock reaching the shell before its collapse.}
\label{fig:PovPsedov}	
\end{figure}
\begin{figure}
\centering
	\includegraphics[width=.47\textwidth]{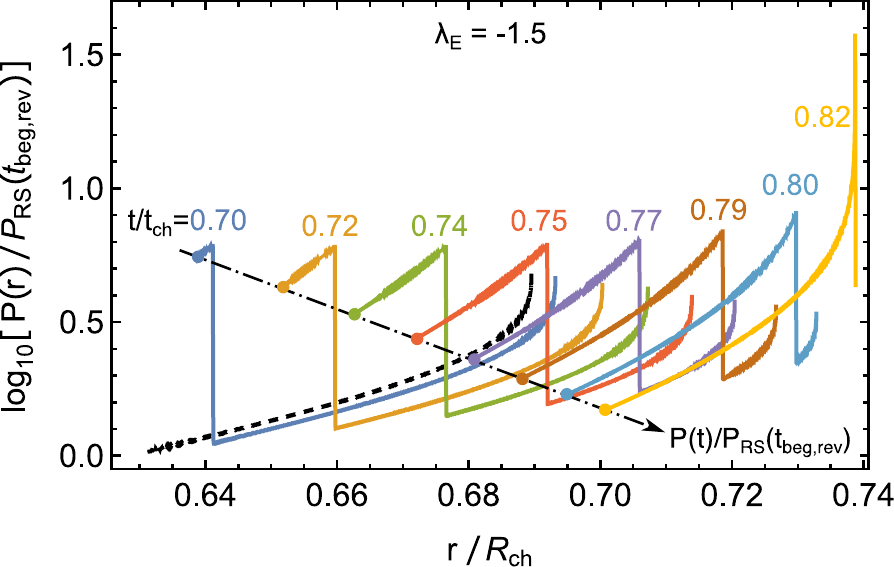}
        \caption{Spatial variation of the pressure scaled with the pressure downstream of the RS at  $\tbegrev$ and close to the shell surface, as obtained from the numerical data for the chosen case of $\lambda_E=-1.5$ (but a similar behaviour is shown by other models).
        Different curves (colors) represent the pressure profile at different times, starting very close to $\tbegrev$ (for this model: $\tbegrev=0.694\,\tch$) and at positions between $\Rshell$ and $R_{\mathrm{CD}}$.
        The temporal variation of the pressure here is highlighted by the dot-dashed line, and dots mark the position from where we extract the profile shown in Fig.\ref{fig:PovPsedov}, bottom panel.
        The black dashed curve represents instead the pressure profile an instant before the beginning of reverberation.
        }
\label{fig:PovPRS_space}	
\end{figure}

A first result comes from the comparison, for some representative values of $\lgEeff$, between the SNR pressure that actually pushes against the shell ($P\rs{outer}$), and the pressure derived at the center of the Sedov solution, namely:
\begin{equation}
   \Psedov(t)=0.0489\,\left(\frac{t}{\tch}\right)^{-6/5}\frac{\rhoism\Esn}{\Mej}\,,
\label{eq:psedov}
\end{equation}
which provides a good approximation even if compared to the pressure scaled with the downstream pressure at the FS (see above). 
From Fig.~\ref{fig:PovPsedov} (upper panel) it can be clearly seen that the ratio between these two pressures is far from being constant in time, and that the trend of this ratio also strongly depends on $\lgEeff$.
These scaled profiles all show a peak right at the beginning of reverberation, then a decrease to much smaller values, typically one order of magnitude, and finally a new strong increase, mostly due to the geometrical fact that the flow is converging.
According to the Guderley solution \citep{Guderley:1942}, the outer pressure should increase like $(\timplo-t)^{-0.6232}\propto R^{-0.9054}$, where $\timplo$ is the time at which the shell implodes.
With this trend near the implosion, a smooth approximation of the pressure near that critical time is possible if the pressure is multiplied by a power of $\Rshell$ close to 1 or larger. In the next section we will consider it multiplied by $4\upi\,\Rshell^2$, so that beyond making the treatment easier it translates into a physically meaningful quantity, namely the force pushing the shell on the outer side.
The lower panel of Fig.~\ref{fig:PovPsedov} instead shows the evolution of $\Pouter(t)$ scaled with the pressure downstream of the RS, computed at $\tbegrev$, given by 
\begin{equation}
P_{\mathrm{RS}}(\tbegrev)=\frac{2}{\Gamma+1}\,\rhoej(t)[\dot{R}_{\mathrm{shell}}(t) - \dot{R}_{\mathrm{RS}}(t)]^2,
\label{eq:pressRS}
\end{equation}
where $\Gamma$ is the adiabatic index (equal to 5/3 in this case). 
The (first) peak of each curve gives the initial pressure jump at the  reflected shock (of a moderate intensity), almost the same factor of $\sim5-6\, \,P_{\mathrm{RS}}(\tbegrev)$ for all models.
Each curve shows an initial sharp decrease of pressure, then followed by a rapid increase at the moment the shell collapses.
The reason for this early decrease can be better appreciated by looking at Fig.~\ref{fig:PovPRS_space}, which displays the normalized pressure profiles at different times close to $\tbegrev$, and at positions ranging between $\Rshell$ and $R_{\mathrm{CD}}$, for the model $\lambda_E=-1.5$. 
One can notice that the strength of the reflected shock is originally constant with time, while it possibly increases when it approaches the CD. 
Due to the positive pressure gradient behind the reflected shock, the pressure at the boundary with the shell (at each time highlighted with a filled circle), instead sharply decreases with time.
Note how the time step between subsequent curves in the plot is quite small, which means that the pressure decrease at the shell  boundary is quite fast.

\subsection{Acceleration and number of shocks reaching the shell}

Let us introduce the acceleration induced by the outer pressure:
\begin{equation}
    \aouter(t)=\frac{4\upi \Rshell^2\Pouter}{\Mshell},
\label{eq:aouter_first}
\end{equation}
namely the ratio between the force exerted from the outer pressure and the shell mass.
This quantity is equal to zero for all times $t<\tbegrev$, has the advantage of vanishing at $\timplo$, and is closely connected to the time evolution of $\Rshell(t)$, being its second derivative in time.
\begin{figure*}
\centering
	\includegraphics[width=.82\textwidth]{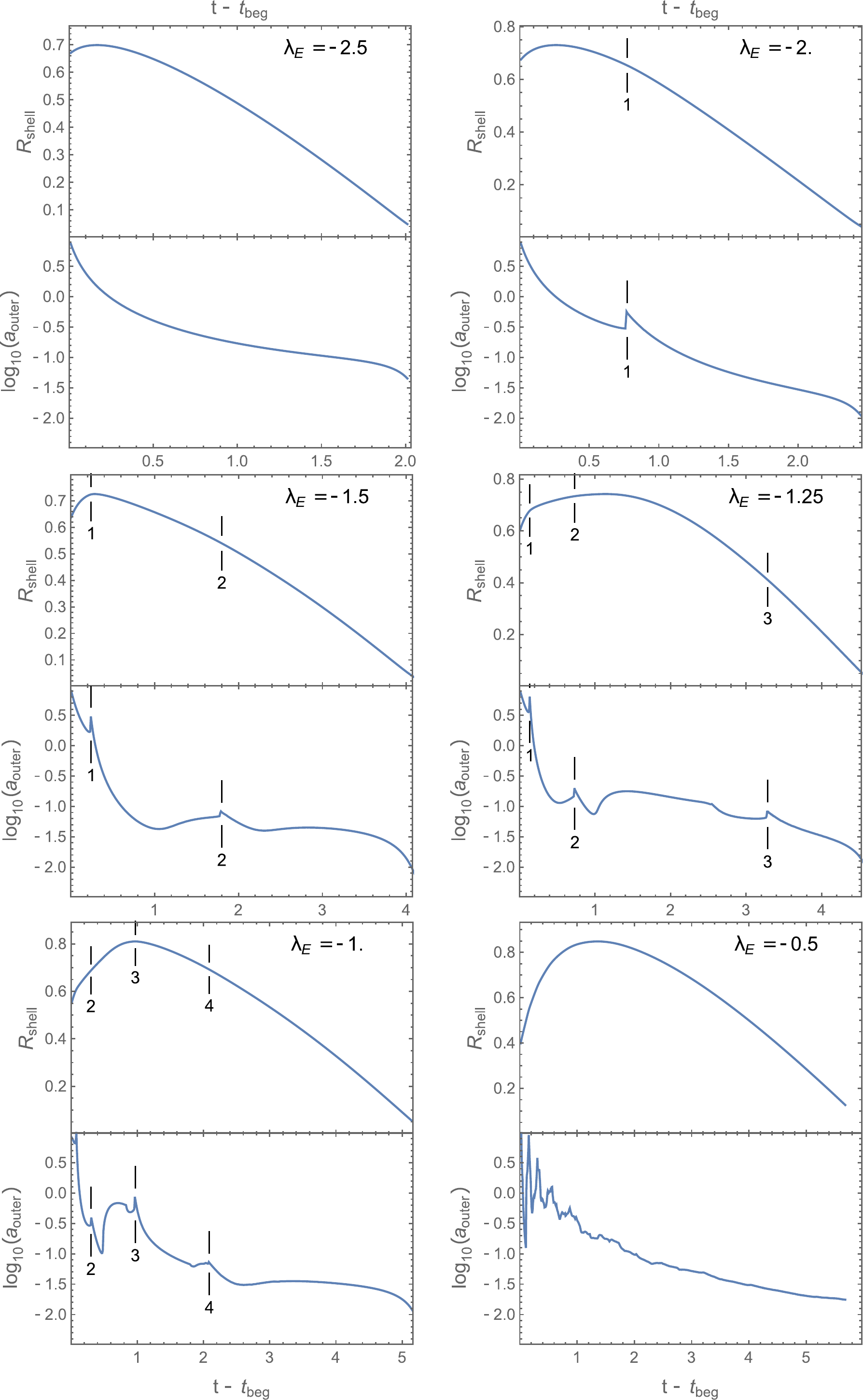}
        \caption{Behavior of $\aouter(t)$ and $\Rshell(t)$ for some representative $\lgEeff$ values (see text for details).}
\label{fig:overviewmodels}	
\end{figure*}
\begin{figure}
\centering
	\includegraphics[width=.42\textwidth]{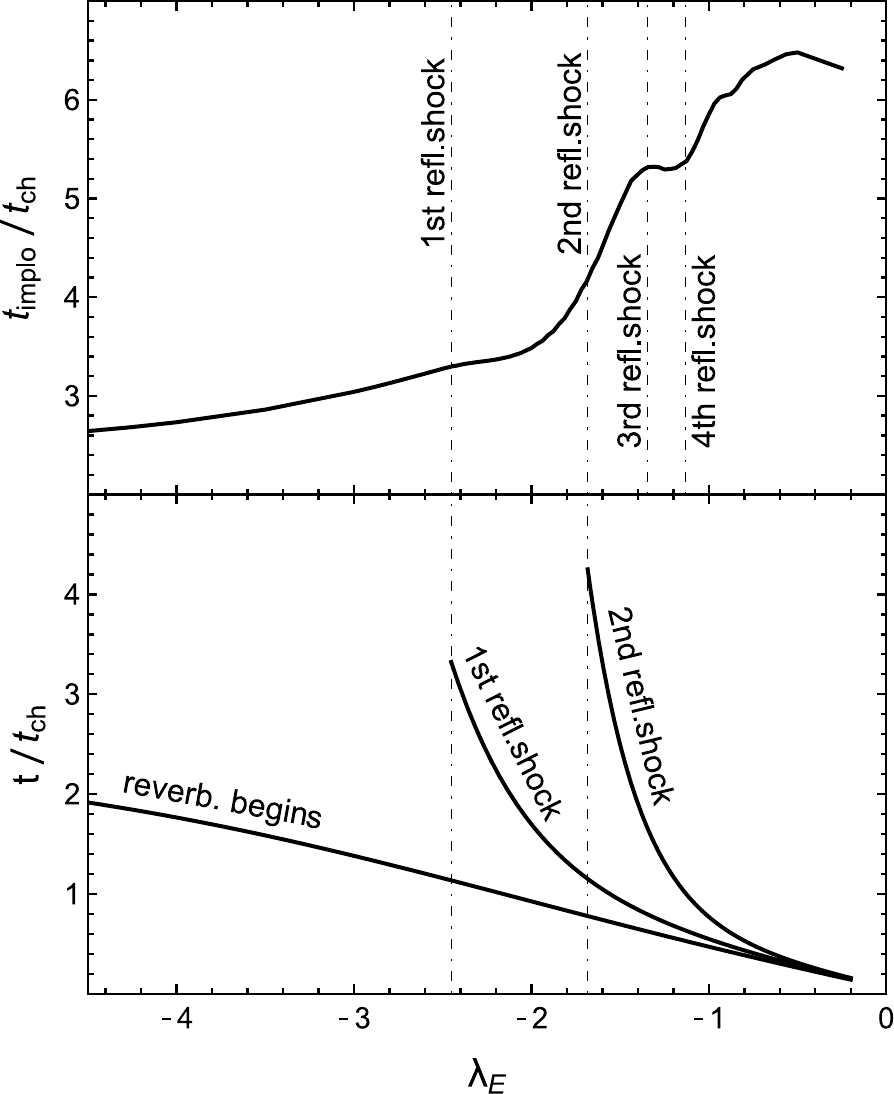}
        \caption{Upper panel: Dependence of the implosion time on the quantity $\lgEeff$, directly related to the mass of the shell. As it is explained in the text, the complex trend shown in this figure is due to the combination of different regimes, essentially depending on the number of reflected shocks that reach the shell before it eventually implodes. Bottom panel: Dependence of $\tbegrev$, $\tfstsh$ and $\tsndsh$ on $\lgEeff$, as given by Eq.~\ref{eq:tbegrev29}-\ref{eq:t2_31}.}
\label{fig:timploCurve}	
\end{figure}
Let us now analyze in more detail the behavior of $\aouter(t)$ over our wide range of $\lgEeff$ values.
As can be seen in Fig.~\ref{fig:overviewmodels}, one may identify different regimes, depending on the number of reflected shocks impinging on the shell before it reaches the center (shown as jumps in $\aouter$).
For $\lgEeff<-2.45$ no reflected shock reaches the shell, so that $\aouter$ shows a smooth behavior for all times between $\tbegrev$ and $\timplo$. 
One reflected shock occurs in the range $-2.45<\lgEeff<-1.68$, while also a second reflected shock hits the shell in the range $-1.68<\lgEeff<-1.35$. 
Around $-1.35$ also a third shock appears, this time preceded by a feature looking like a rarefaction wave; also a fourth shock, appearing at $\lgEeff$ higher than $-1.13$ shows a qualitatively similar pattern.
At higher $\lgEeff$ values the complexity of the pattern of reflected shocks increases dramatically.
Incidentally, $\lgEeff=-2.20$ is the transition value between the cases in which the shell hits a still expanding RS, or an already contracting one.
 
In Fig.~\ref{fig:overviewmodels} one may notice some trends: first, the strength of the jump associated to a given shock is getting higher at higher $\lgEeff$, namely when the shock reaches earlier the shell.
This can be easily explained: when a shock reaches the shell soon after $\tbegrev$ the shell is still expanding, or slowly receding, so that the relative velocity is higher and the dynamical effect stronger; on the opposite, at very late times the shell recession is more prominent, and then the relative velocity lower.
Moreover, a jump in the pressure, namely in the second time derivative of the shell radius, does not have an immediate effect on its radial evolution.
In particular, when the number of shock impacts becomes very large (say for $\lgEeff\simeq0.9$ and above), their cumulative effect on the evolution of $\Rshell$ is not much different from that of a more regular pressure evolution.
Therefore the strongest effect of the individual shocks on the $\Rshell$ evolution is in a $\lgEeff$ range around $-1.0$, when only a few major shocks are involved.

\subsection{Fittings for the beginning of reverberation and the arrival of shocks to the shell}

The following formulae give accurate fits to 
$\tbegrev$, as well as to the arrival times of the first two reflected shocks:
\begin{eqnarray}
    \tbegrev(\lgEeff)& \simeq & 2.4102 \; \frac{1-\exp(-0.1494+1.1606\,\lgEeff)}{1+\exp(1.6831 + 0.6805\,\lgEeff)}\,\,\tch, \hspace{1cm} \label{eq:tbegrev29}\\
    \tfstsh(\lgEeff) & \simeq & \tbegrev(\lgEeff)+\exp(-4.8640-2.3026\,\lgEeff)\,\,\tch,\hspace{0.5cm} \label{eq:t1_30} \\
    \tsndsh(\lgEeff) & \simeq & \tbegrev(\lgEeff)+\exp(-4.8023-3.5932\,\lgEeff)\,\,\tch.\hspace{0.5cm} \label{eq:t2_31}
\end{eqnarray}
%
The differences between the arrival times of the various shocks and $\tbegrev$ are well approximated as an exponential of the (logarithmic) quantity $\lgEeff$, which means that they are approximate power laws of $\LztauzEFF$.
These approximations have an accuracy with respect to the lagrangian models of about $0.01\,\tch$, and of course they are valid only for the range of $\lgEeff$ in which the relative shock is present.
The time of the begining of reverberation and the arrival times of the first two reflected shocks are shown in 
Fig.~\ref{fig:timploCurve}.

As for $\timplo$, since its value is affected by the pressure evolution associated to the arrival of the various reflected shocks, it shows a very regular behavior only for $\lgEeff<-2.45$, while for higher values its behavior is much more structured, as also shown in Fig.~\ref{fig:timploCurve}.
\subsection{Fittings for the shell radius}

It should be clear that the evolution of the shell is too complex to be represented by a simple analytical solution, so that we search for approximate fitting formulae.
In the following we introduce some interpolating functions, whose structure does not intend to have any specific physical meaning, but only to provide reasonably accurate interpolations by limiting as much as possible the number of free parameters.
According to what we explained above, while we aim at approximating $\aouter$, associated to the outer pressure, we do not really need to match its pattern in detail, a particularly complex task especially in the presence of one or more reflected shocks.
Instead, it is more important to match the dynamical ``effects'' of this pressure, and with the strategy described above a smoothed version of the evolution of $\aouter$ is effectively obtained.
We have found that, for most $\lgEeff$ values, excellent fits to the evolution of $\Rshell$ can be obtained with a function whose second derivative is given by the product of a second order polynomial times a decreasing exponential, namely:
\begin{equation}
\label{eq:aouter}
    \aouter(x)=(1-a x+b x^2)\,e^{c-k x}\frac{\Rch}{\tch^2},
\end{equation}
where $x=(t-\tbegrev)/\tch$.
The corresponding formula for $\Rshell$ is then:
\begin{eqnarray}
    \Rshell(x)&=&\tch^2\int_0^x\,dx'{\int_0^{x'}\,dx''\,\aouter(x'')}\;=\nonumber\\
    &&\!\!\!\!\!\!\!\!\!\!\!\!\!\!\!\!\!\!\!\!\!\!\!\!\!\!\!
    \frac{e^c}{k^2}\left\{\left[1-e^{-k x}-k x\right]-\frac{a}{k}\left[2\,(1-e^{-k x})-(1+e^{-k x})k x\right]\right.\nonumber\\
    &&\!\!\!\!\!\!\!\!\!\!\!\!\!\!\!\!\!\!\!\!\!\!\!\!\!\!\!
    \left.+\frac{b}{k^2}\left[6\,(1-e^{-k x})-2\,(1+2e^{-k x})k x-e^{-k x}k^2x^2)\right] \right\}\Rch\nonumber\\
    &&\!\!\!\!\!\!\!\!\!\!\!\!\!\!\!\!\!\!\!\!\!\!\!\!\!\!\!
    + \; (\Rbegrev+\Vbegrev\tch x),
\label{eq:Rshell_fit}
\end{eqnarray}
where the quantities $\Rbegrev$ and $\Vbegrev$ are not free parameters, but are derived from the pre-reverberation evolution.
The 4 parameters $a$, $b$, $c$, and $k$ are derived, for each value of $\lgEeff$, from a best fit to the numerical evolution of $\Rshell$.
For $\lgEeff\lesssim-1.8$ this functional form allows fits with residuals always less than 0.003; while worse fits are obtained for $\lgEeff\gtrsim-1.3$, with residuals anyway not exceeding 0.03.
For this reason we have decided to adopt the same functional form for all the $\lgEeff$ considered.

\begin{figure}
\centering
	\includegraphics[width=0.42\textwidth]{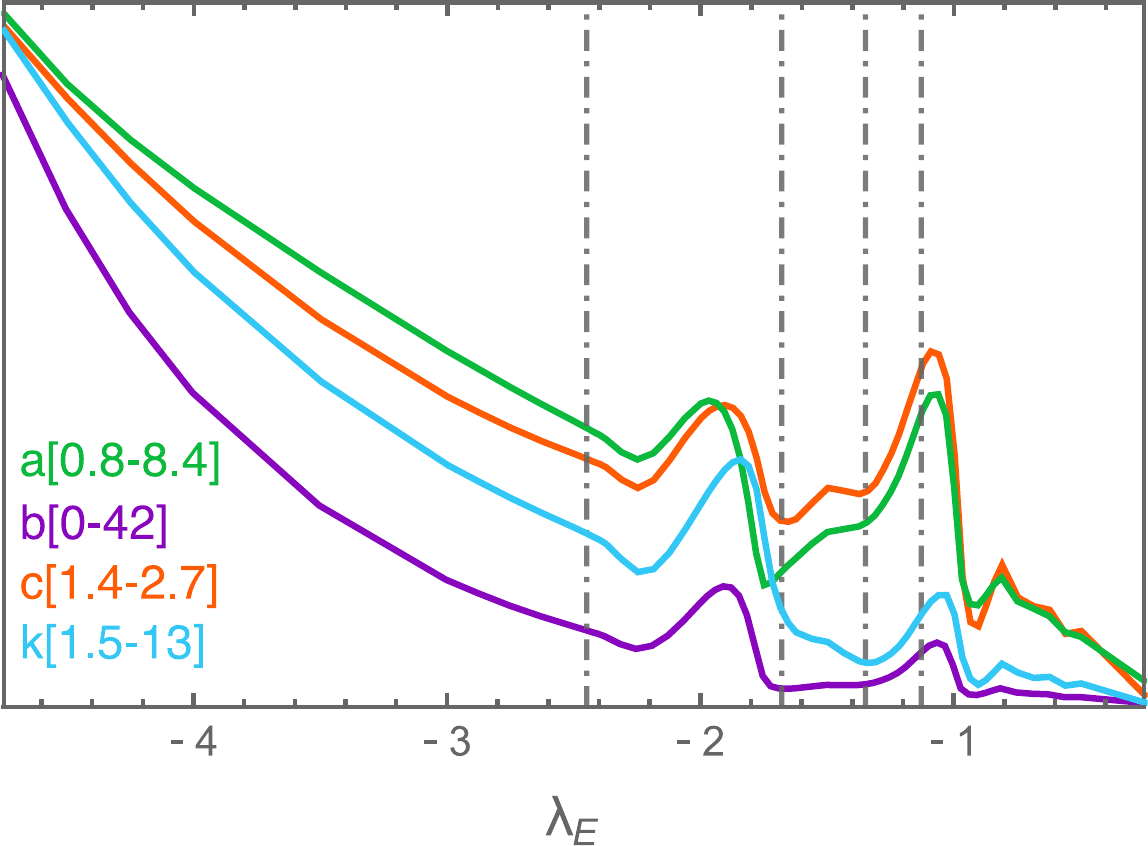}
        \caption{Dependence on $\lgEeff$ of the parameters $a$, $b$, $c$, $k$, used for the fits. Since the scope of the figure is to show the appearance of different regimes with varying $\lgEeff$, we have preferred to condense the curves on the same plot, even if their vertical scale is different. The ranges of their vertical axes are then given in brackets in the legend. Different colors correspond to different parameters. The vertical dot-dashed lines indicate the positions of the first, second, third, and fourth reflected shocks.}
\label{fig:parameterswholerange}	
\end{figure}
The trends of the best-fit parameters are shown in Fig.~\ref{fig:parameterswholerange}.
It is apparent that the trends are regular only when $\lgEeff<-2.45$ (no reflected shock).
Analytic approximations for the functional dependence of the parameters over $\lgEeff$ are given in Appendix~\ref{sec:acc-formula}.
The formulae that we have devised are quite complex, because we had to add extra components to the overall trend, in order to reproduce all the features that appear in Fig.~\ref{fig:parameterswholerange}.
A high accuracy is required, and with our fits we have reached residuals with mean squared deviations $\simeq0.01$, because the evolution of $\Rshell$ is strongly dependent on the values of these parameters.
In Appendix~\ref{sec:acc-formula} we also give a simpler fit to $\aouter$, valid for very low energy models ($\lgEeff<-4.5$).
In Sec.~\ref{sec:rev-thin-shell} we will show how to apply these formulae to perform an approximate modelling of the PWN+SNR evolution, with our modified thin-shell approach.

\begin{figure}
\centering
	\includegraphics[width=.42\textwidth]{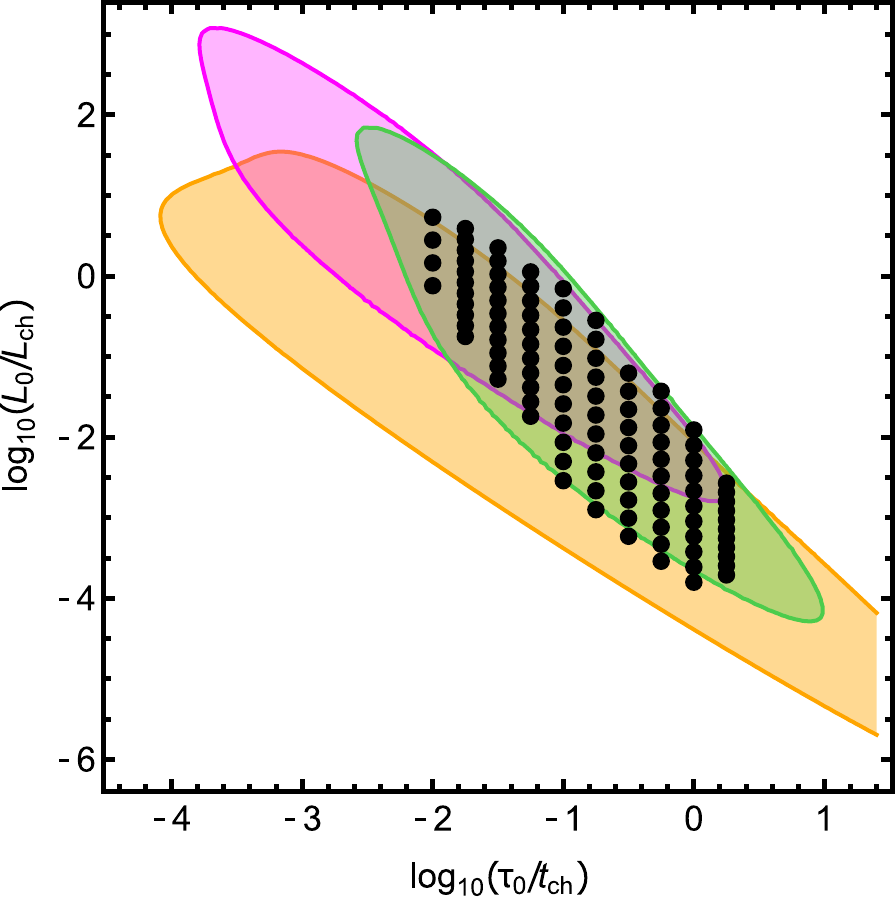}
        \caption{This figure displays the same region of the parameter plane as in Fig.~\ref{fig:moredistributions}, but we the aim of showing, with black dots, the positions in the parameter plane of all the lagrangian models we have run. The filled areas show the regions with 95\% of the elements for each synthesized population, the colors of which correspond to those in Fig.~\ref{fig:moredistributions}.
        }
\label{fig:regionsandmodels}	
\end{figure}
\begin{figure}
\centering
	\includegraphics[width=.47\textwidth]{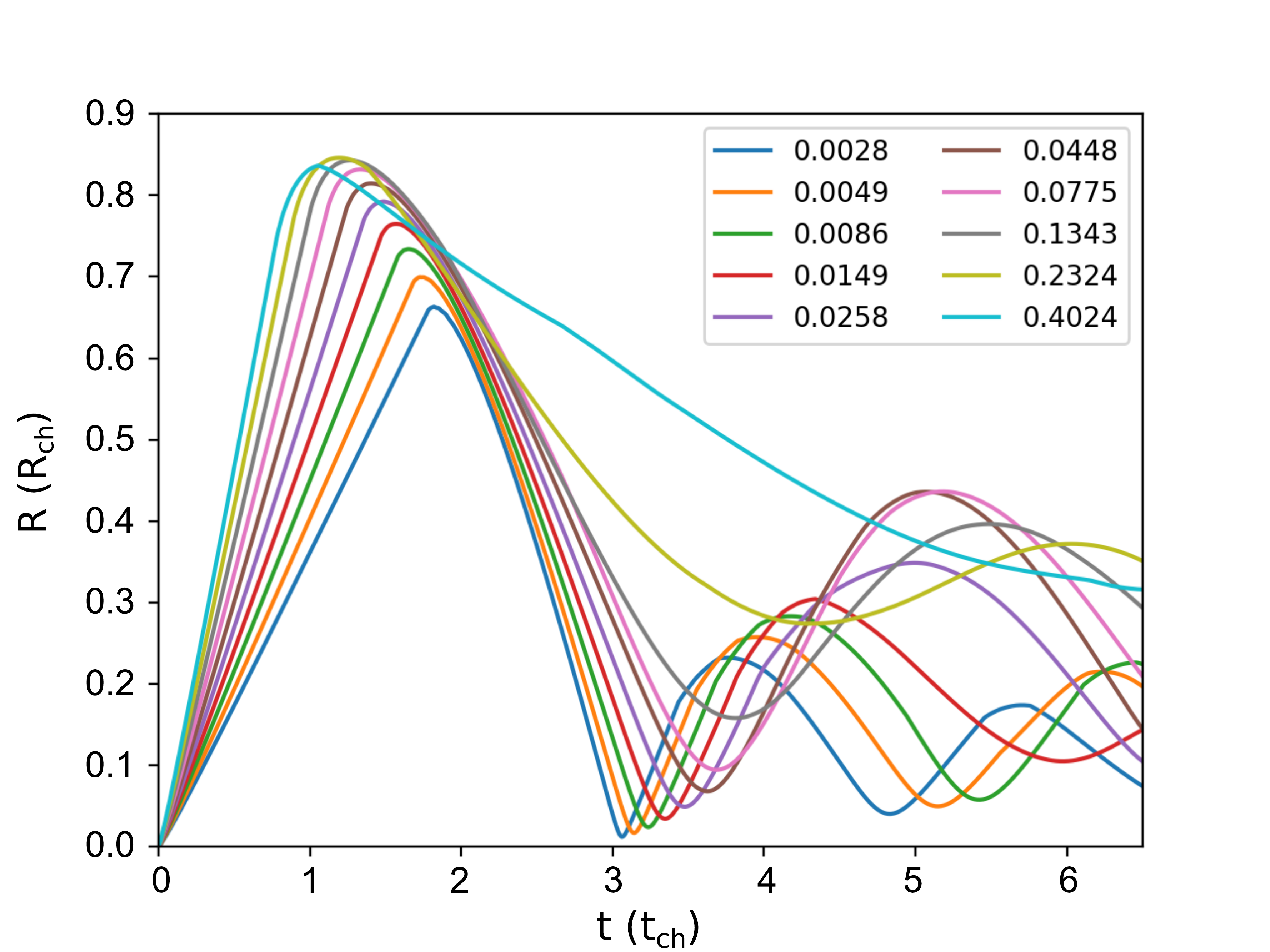}	
	\vspace{-0.3cm}
    \caption{Evolution in time of the PWN radius for pulsars with $\tauz=0.1\tch$, and different values of $\Lz/\Lch$ (given in the legend).
    In comparison with Fig.~\ref{fig:moredistributions}, this plot represents models with $\lgLz$ varying within the green area and fixed $\lgtauz=-1$. This was chosen as a representative case, being almost centered in the PWNe distribution.}
\label{fig:pwnevol}	
\end{figure}
\section{A PWN interacting with the SNR}
\label{sec:pwn}
%

\subsection{Dynamical evolution}

In this section we first present the results of about one hundred numerical simulations of the coupled evolution for a PWN+SNR system, obtained using our lagrangian code, and ranging from very early phases ($t\ll \tauz$) until rather late stages of the reverberation phase.
The location of these models in the $\lgtauz)$\,--\,$\lgLz$ parameter plane is shown in Fig.~\ref{fig:regionsandmodels}.
They are equally spaced in $\lgtauz$, in the range $[-2.0,\,0.25]$ with step size of 0.25.
The ranges of $\Lz$ have been chosen to reasonably cover areas in the parameter plane with higher density of pulsars, using a compromise between distributions in the literature (see Sec.~\ref{sec:pplane}).

The models assume the PWN radius, $\Rpwn$, as the inner boundary of the lagrangian grid. Initial conditions for the simulations are given in Appendix~\ref{sec:Vearly-pwn-evoGEN}.
In order to carefully resolve the structure of the swept-up shell that forms during the evolution, and in particular to correctly model its dynamics during the reverberation phase, we have subdivided the spatially uniform ejecta between $\Rpwn$ and $\Rsnr$ in 5000 or more equally spaced shells.
The resolution in the outer ISM is less important, so we opted for shells spaced of about $0.1\U{ly}$.
The maximum size of the grid is chosen to prevent the FS from reaching the grid boundary along all the calculation, always lasting more than $6\,\tch$.

As a reference, in Fig.~\ref{fig:pwnevol} we show the evolution in time of $\Rpwn$, for $\tauz/\tch =0.1$, and different values of $\Lz/\Lch$.
It is evident that, as $\Lz/\Lch$ increases, the first maximum of $\Rpwn$ is reached at earlier times, even though after the beginning of the reverberation phase the massive shell continues expanding for a longer time; while after then the asymptotic velocity of the PWN contraction is lower, also because more mass has been collected in massive shell, and consequently the time of the first PWN maximum compression is delayed.
Then the nebula re-expands, experiencing a series of less severe expansions and compressions, as already found in \citet{van-der-Swaluw:2001} and \citet{Blondin:2001}.
Note that in the lower-luminosity regime the smallest size of the PWN is reached during the first compression; as the luminosity of the pulsar rises, the evolution of the PWN radius becomes more complex, and the first compression might be no more the strongest one.
At higher energies the dynamics is complicated by the presence of several weak internal shocks  at $\tbegrev$, when the swept-up shell meets the RS, and that keep propagating and bouncing between the CD and the PWN radius, as already noticed in the case of the shell+SNR interaction (see Sec.~\ref{sec:shell}).

\begin{figure}
\centering
	\includegraphics[width=.5\textwidth,viewport=95 5 360 220,clip]{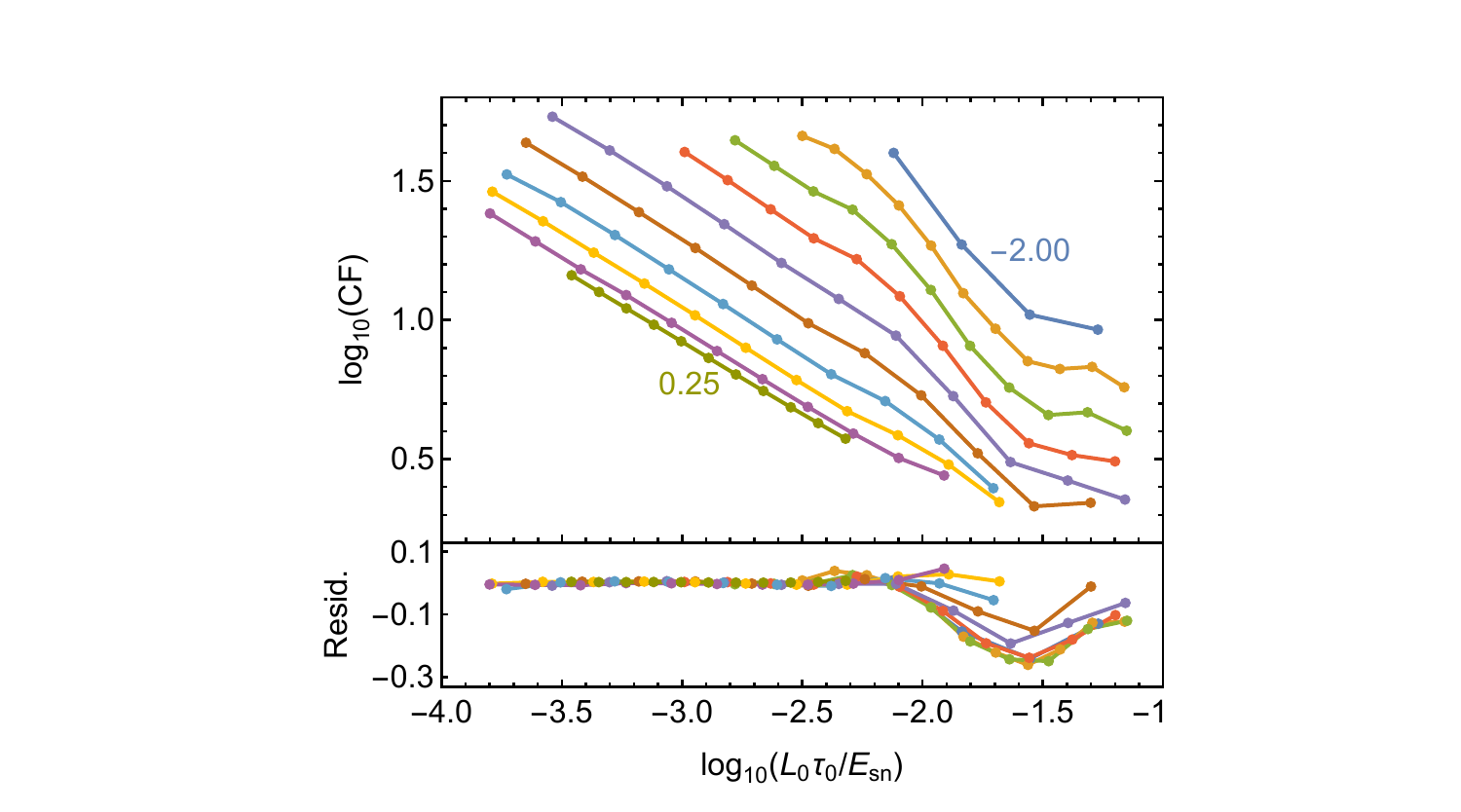}
    \vspace{-0.3cm}
    \caption{The upper panel gives $\lgCF$ as a function of $\Lz/\Lch$, computed for all our models. Each line refers to a given value of $\tauz$, starting from $-2.00$ (top right) to $0.25$ (bottom left), and each dot represents a single model. It is apparent that, for $\lgLztauz\lesssim-2.5$ the trend are almost linear. In the text we give an approximating function for this region, and the lower panel displays the residuals after subtracting this function. In the higher energy regime, the most noticeable displacements are: one  small bump in the positive direction, at $\lgLztauz\sim-2.4$ (actually, near to the value above the first reflected shock reaches the PWN; and another, more prominent, in the negative direction around $-1.6$ (near the value above which also a second reflected shock appears).
}
\label{fig:CFinLagrangianMdls}	
\end{figure}

\subsection{Compression factors}

\begin{figure}
\centering
	\includegraphics[width=.47\textwidth]{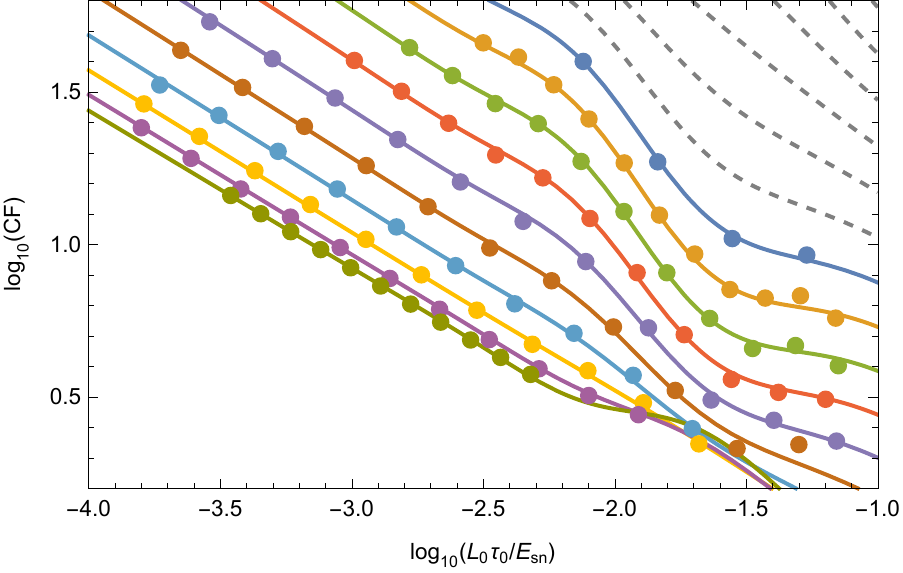}
    \vspace{-0.3cm}
    \caption{Analytic approximation (formula given in the text) to the CF compared to the numerical values from lagrangian models (shown by dots). The gray dashed lines represent the extrapolation to $\lgtauz<-2.0$, in steps of 0.25.
}
\label{fig:CFgoodapprox}	
\end{figure}
Let us now focus on the behavior of the CF: the 
trend of this quantity, for all our models, is shown in Fig.~\ref{fig:CFinLagrangianMdls}.
For low $\lgLztauz\lesssim-2.5$ one may derive an approximation function (with an average error of about 1\% with respect to numerical results), linear in $\lgLztauz$: 
\begin{eqnarray}
    \lgCFlow\!\!\!\!&=&\!\!\!\!-0.6079-\frac{0.1986}{0.4117+(\tauz/\tch)}\lgtauzD
    \nonumber\\
    &&\!\!\!\!\!\!\!\!\!\!\!\!\!\!\!\!
    -\left(0.5247-0.0294\lgtauzD\right)\lgLztauzD.
    \label{eq:CFlow}
\end{eqnarray}
%
An extension of this formula to higher energies can be done by fitting the residuals shown in the lower panel of Fig.~\ref{fig:CFinLagrangianMdls}.
A suitable expression consists in the sum of two Gaussian functions in $\lgLztauz$, whose magnitudes are only functions of $\tauz$, namely:
\begin{eqnarray}\label{eq:CFfull}
    \lgCF\!\!\!\!\!&=&\!\!\!\!\!\lgCFlow+f_1(\tauz)\exp\left(-6.4\,(2.07+\lgLztauzD^2\right)\nonumber\\
   &&\!\!\!\!\!
   -f_2(\tauz)\exp\left(-3.25\,(1.77+\lgLztauzD^2\right)\,,
\end{eqnarray}   
where:  
\begin{eqnarray}\label{eq:CFfull_par}
   f_1(\tauz)\!=&&\!\!\!\!\!\!\!\!\!\!\!\!\!1.724\!+\!0.558\log_{10}\!\left(\frac{\tauz}{\tch}\right)\!-\!
    \log_{10}\!\left[1\!+\!59.70\left(\frac{\tauz}{\tch}\right)^{0.815}\right]\!,    \nonumber\\
    f_2(\tauz)\!=&&\!\!\!\!\!\!\!\!\!\!\!\!\!1.868\!+\!0.673\log_{10}\!\left(\frac{\tauz}{\tch}\right)\!-\!
    \log_{10}\!\left[1\!+\!94.41\left(\frac{\tauz}{\tch}\right)^{1.063}\right]\!.
\end{eqnarray}
Fig.~\ref{fig:CFgoodapprox} shows the result of this approximation, compared to the numerical values, obtained with our lagrangian models.
The fit is excellent (average error below 3\%), and the extrapolation to smaller values of $\tauz$, the most interesting ones because they are associated to higher CF values, looks rather smooth.
However, we warn that such extrapolation is not physically grounded, so it is not clear how far it could be extended.
One may only speculate that the low-energy trend of the CF, being so simple and regular, should reasonably extend to even lower energies than those actually tested.
On the other hand, cases with $\lgtauz<-2.0$ are very demanding from the numerical point of view since now, in order to keep $\tini\ll\tauz$, very small initial times, and then smaller time-steps are required.
This also means that this range is extremely sensitive to the initial conditions, since almost the entire energetics is released in the very first phase of evolution. 

In a general, qualitative sense, one may notice that the CF increases when decreasing the energetics of the pulsar, as well as when decreasing $\tauz$, so that the most compressive cases are in the lower leftmost side of Fig.~\ref{fig:regionsandmodels}.
To our purposes they represent the most interesting cases, especially in the perspective of a future analysis of the effect of radiative losses on the PWN evolution.

\subsection{A reference model}

\begin{figure}
\centering
	\includegraphics[width=.47\textwidth]{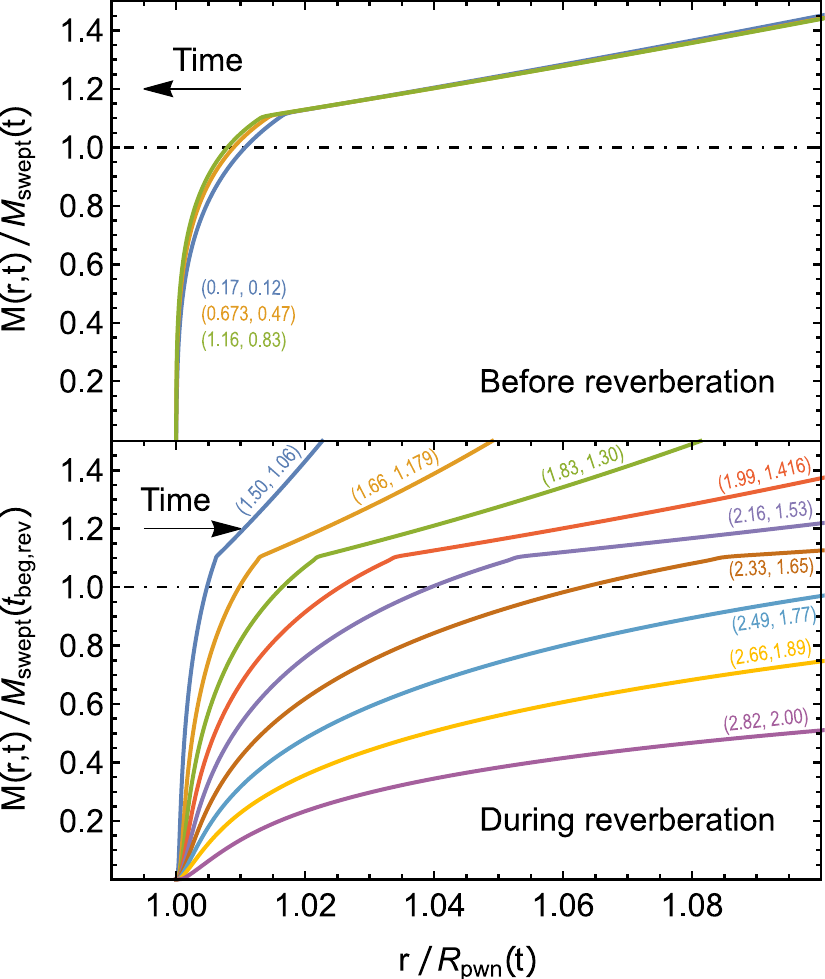}
    \vspace{-0.3cm}
    \caption{Time behavior of the mass radial profile, for our reference model. The radial coordinate is scaled with the PWN radius, while the enclosed mass is scaled with the swept-up mass from the thin-shell model for the pre-reverberation evolution. Different colors are used for different times given in terms of ($t/\tch, t/\tbegrev$), and the arrows show in which direction the profile evolves with increasing time. For this figure we have used numerical data from a reference model $\lgLz=-2.0634$, $\lgtauz=-1.0$, but the results are qualitatively independent from which model parameters are used.} 
\label{fig:MofRscaled}	
\end{figure}
Let us now investigate in more detail one of our lagrangian models, with $\lgLz=-2.0634$ and $\lgtauz=-1.0$.
The quantitative results that are found here are then specific for this case, but nonetheless they are representative of aspects that are qualitatively rather general in our models.

\noindent First, let us discuss the assumption that the mass accretion on the shell stops with the beginning of reverberation: this recipe was introduced already by \citet{Reynolds:1984}, but it has not been used in most of the literature on the subject. 
During the pre-reverberation phase (and most effectively at earlier times $t\sim\tauz$) the shell was collecting mass from the outer, unshocked ejecta (see Sec.~\ref{sec:pre-revPWN}).
As already mentioned in Sec.~\ref{sec:shell}, at that time a shell is characterized by a low relative velocity with respect to the cold, unshocked ejecta; thus the entropy of the shocked medium is low, and therefore the downstream material can be effectively compressed, i.e.\ it can effectively stick to this shell.
During reverberation, instead, that shock is moving with a much higher relative velocity with respect to the hot, already shocked ejecta; therefore, in this case the downstream medium has a much higher entropy and cannot be easily compressed.
This point was also discussed in our Paper I.

Such behavior has been addressed more quantitatively in Fig.~\ref{fig:MofRscaled}.
The meaning of the individual profiles can be better appreciated by having in mind that all the mass is external to the PWN (i.e.\ $M(\Rpwn(t))=0$), and that regions with a steeper profile mean a higher density there.
In the upper panel the radial profile of the enclosed mass is plotted, for several times earlier than $\tbegrev$.
The radial coordinate is in units of $\Rpwn(t)$, while the mass is in units of the swept-up mass, computed as the mass of the ejecta that, in the absence of the PWN, would be enclosed in the sphere with radius $\Rpwn(t)$.
These profiles, for a scaled radius ranging between 1 and about 1.02, are directly related to the density profile inside the shell, and the fact that they are almost superimposed means that the density preserves its profile, apart from a slight decrease with time of the shell relative width.
The sharp break in the profiles (reflecting a density jump) indicates the position of the shock at the outer boundary of the shell; while the scaled mass slightly higher than unity means that the real swept-up mass is that within the outer boundary of the shell, rather than within $\Rpwn(t)$.
After $\tbegrev$ (lower panel) the evolution changes completely: the mass within the shell, now scaled with the swept-up mass at $\tbegrev$, does not change with time, as it can be inferred from the constancy of the vertical coordinate of the break; on the other hand, the relative width of the shell quickly increases with time, partly reflecting its physical broadening, and partly as a consequence of a decreasing of its size.
This figure then justifies the assumption of a fixed shell mass during reverberation and, in addition, it shows that, when the PWN has been compressed, the needed conditions for treating the shell as a thin-shell may no longer be valid.
%

\begin{figure}
\centering
	\includegraphics[width=.44\textwidth]{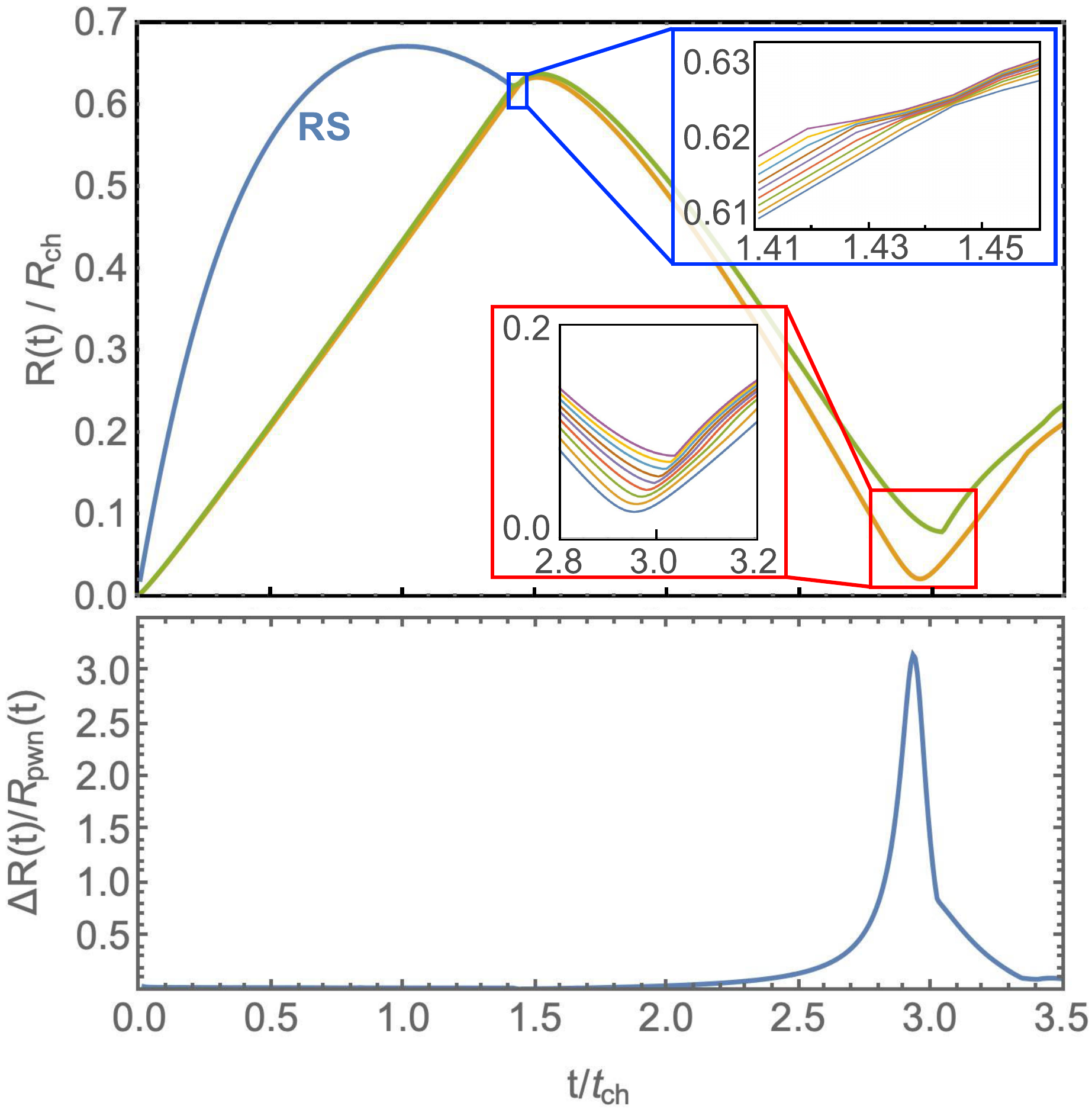}
    \caption{Upper panel: Time evolution of the inner (orange color -- corresponding to the PWN) and outer (green color) radius of the shell, as computed with our lagrangian code, for our reference model.
    Two insets show the inner structure of the shell, near two specific times: $\tbegrev$ (blue box), and the time of maximum compression (red box).
    The inner lines are curves of equal mass (arbitrarily chosen values).
    Bottom panel: time variation of the shell thickness with time.}
\label{fig:non-thinS}	
\end{figure}
Fig.~\ref{fig:non-thinS} shows, in addition to the general evolution of the shell, two insets displaying the evolution of the structure inside the shell near two topical times: $\tbegrev$ and the time at which the PWN experiences it maximum compression.
For $t<\tbegrev$ the shell boundaries are very close each other, meaning that the thin-shell approximation is well satisfied.
Afterwards, during the reverberation phase, the outer edge of the shell is defined by the mass collected before $\tbegrev$, and one may clearly see that the shell becomes thicker, and as the PWN starts to contract the shell inflates progressively.
Especially close to the maximum compression, the combination of a higher shell thickness and a smaller shell size implies that a thin-shell approach is no longer justified. 

The hydrodynamic effects of a finite thickness can be outlined as follows.
The impact with the RS first, and the push from the PWN at a later phase, generate transmitted and reflected secondary shocks, so that the shell does not fully behave as a rigid body (as clearly shown in Fig.~\ref{fig:non-thinS}).
In addition, the travelling of these secondary shocks across the shell contributes to convert part of the mechanical work on the shell into internal energy, something that would not happen in an infinitely thin shell.

\section{A revised thin-shell model}
\label{sec:rev-thin-shell}

The lagrangian models presented in the previous section allow us a detailed investigation of the dynamical behavior of the swept-up shell.
These results can be compared with those from a thin-shell approximation, both before and during the reverberation phase.
In principle, the results of the numerical models are more accurate than the corresponding thin-shell approximation, and therefore can be used to test the level of accuracy of the thin-shell approach, in its various forms.
On the other hand, a weakness of the numerical models derives from the need of setting a finite, and not exceedingly small, value for $\tini$ (we have typically used $\tini\sim 0.03$--$0.04\,\tauz$) and, in addition, from the large computing time required to perform a finely spaced coverage in the parameter plane.

The goal of our modified thin-shell model, rather than just reproducing results from the lagrangian simulations, is indeed to extend the range of predictions mainly in the two following directions:\\
\noindent(i) Cases with very small $\tauz/\tch$,
    where lagrangian simulations are not efficient due to the requirement of very small initial times, and lead to lower accurate results due to the extreme pressure contrast at the two sides of the (now no longer infinitely thin) mass shell.
\noindent(ii) Cases accounting for dynamical effects of radiative losses in the PWN; an advantage of the thin-shell approach is to be enough computationally light to allow a treatment of synchrotron + inverse Compton losses, of their effect to modify the energy distribution of the relativistic particles and ultimately the pressure associated with these relativistic particles.
Unfortunately, coupling 1D hydrodynamic models with a routine that evolves the particle energy distribution would be too heavy to be run in more than just very few cases.

\subsection{Comparing numerical and semi-analytical results}

We have first compared numerical (i.e.\ lagrangian) and semi-analytic (i.e.\ using Eqs.~\ref{eq:magfluxcons} and \ref{eq:momentumcons}) solutions in the pre-reverberation phase.
We have then found that the ratios of corresponding quantities, like $\Rpwn(t)$ or $\Qpwn(t)$, converge rather soon but not exactly to unity.
Typically we have noticed mismatches $\sim4\%$ for $R$ and $\sim6\%$ for $Q$.
This could either be due to the thin-shell approximation, or to the choice of a finite $\tini$ in the simulations.
Since running lagrangian models of a PWN+SNR system with an exceedingly small $\tini$ value leads to problems both of computing time and of accuracy, we have then chosen the other way round, by running for this comparison the thin-shell calculation with the same initial conditions as those used for the lagrangian model (see Appendix~\ref{sec:Vearly-pwn-evoGEN}).
In this way the results of the two approaches differ by no more than  $0.2\%$, proving that the thin-shell approach gives quite accurate results, but also that they are more sensitive than expected to the choice of the initial conditions, since a rather long time is required before memory of them is washed out completely.

Even though the following comparative analysis of our thin-shell models with the lagrangian ones has been performed using fully compatible initial conditions, for future uses of our thin-shell model we recommend initial conditions compatible with the analytic expansions as given by Eqs.~\ref{eq:expfirstR} and \ref{eq:expfirstQ}.

In Sec.~\ref{sec:pre-revPWN} we have shown that thin-shell models may allow an accurate description of the pre-reverberation evolution, then we can use them to obtain the physical conditions at the beginning of the reverberation phase.
To compute the evolution in the reverberation phase, one needs first to evaluate $\tbegrev$, as the time at which the PWN and the RS (given by Eq.~\ref{eq:RRS}) intercept. An easier and still accurate way is to use the PWN radius as approximated by Eq.~\ref{eq:Rapprox}. Then one has to evaluate at $\tbegrev$ the following quantities: the shell radius $\Rshell(\tbegrev)$, its velocity $\Vshell(\tbegrev)$, its mass $\Mshell(\tbegrev)$, and finally the PWN pressure-related quantity $\Qpwn(\tbegrev)$.
From $\Mshell(\tbegrev)$ one can then derive $\lgEeff$, by using Eqs.~\ref{eq:forLambda}-\ref{eq:forLambdaL}, and then refer to the shell+SNR models (Sec.~\ref{sec:shell}) to obtain an estimate of the outer pressure.
Notice however that Eq.~\ref{eq:tbegrev29} is strictly valid only for the case of a shell interacting with the SNR, while it  only roughly approximates the case of a PWN interacting with the SNR; the two are almost equivalent only in the case $\tau_0 \ll \tch$.
For the evolution beyond $\tbegrev$ we will use the results of Sec.~\ref{sec:shell} to model the outer pressure from the shocked SNR medium, with a much higher accuracy than done before.
In this way it is possible to accurately reproduce the first phase of reverberation, at least for all cases in which $\tauz\ll\tch$, until the PWN pressure becomes important again, and the PWN completes its first compression.

The underlying idea is that the massive shell behaves like a ``buffer'': right after $\tbegrev$ the outer pressure has to cope mainly with the inertia of the expanding shell, while the inner push from the PWN has weakened consistently (at times larger than $\tauz$, during the pre-reverberation phase, the PWN pressure decreases approximately $\propto t^{-4}$).
Only after the outer pressure has succeeded halting the shell expansion and reverting it to a compression, the inner pressure starts
to increase again, until it eventually dominates the bouncing-back phase. 
During this last phase the PWN pressure increases enough to counteract the inertia of the now imploding shell, while the effect of the outer pressure has become negligible. 
This evolution could be outlined, in terms of a thin-shell approach, through the following equations:
\begin{eqnarray}
\label{eq:magfluxconsB}
\frac{dQ}{dt}&=&L(t)\,R(t)\,,\\
\label{eq:momentumconsB}
\frac{d^2R(t)}{dt^2}&=&\frac{Q(t)}{\Mshell R^2(t)}-\aouter(t-\tbegrev)\,. \end{eqnarray}
The latter equation describes the evolution of the shell momentum, where typically the inner and outer forces, respectively $Q/R^2$ and $\Mshell\aouter$, are effective at different times.
Even if in its first phase the PWN pressure is negligible, one must keep trace of the evolution of the quantity $Q(t)$ (Eq.~\ref{eq:magfluxconsB}), because its value will then rule the PWN pressure around the time of maximum compression.
An underlying assumption in Eq.~\ref{eq:momentumconsB} is that the shell mass keeps constant during this phase, which in the previous section we have found to be a tenable expectation. Of course, this ultimately relies on how the shell boundaries are defined; but, as already shown in Sec.~\ref{sec:shell}, one cannot assume at the same time that the shell mass does not change and that the outside SNR behaves dynamically unaffected by the presence of this shell.
In the present treatment, after evaluating the corresponding $\lgEeff$, we approximate the outer pressure $P\rs{outer}$ by using the formulae derived in Sec.~\ref{sec:shell}, and explicitly given in Appendix~\ref{sec:acc-formula}. 
During this phase, apart from the fact that the shell has a small but finite width, the system resembles the case shell+SNR investigated in Sec.~\ref{sec:shell}, and this will be the case as long as the internal pressure, from the PWN, will slow down and revert the converging motion of the shell.
Only from that time on the actual SNR pressure will become significantly higher than how modeled here, because the later shell expansion will drive another compression wave / reflected shock into the SNR material, increasing $P\rs{outer}$ even more. This further evolution is however beyond our scope, also because our primary objective is to derive the CF, thus we just need to follow, with a reasonable accuracy, the evolution till a time slightly larger that that of the first maximum compression of the PWN.
A major difference between the lagrangian models and our thin-shell model is that by construction it cannot reproduce oscillations following the first one.
On the other hand, thin-shell models using an outer pressure scaled with the Sedov solution may show oscillations, but in a quantitatively unreliable way.

At a higher level of accuracy there are some additional differences between the shell+SNR and the PWN+SNR problem.
The most relevant one is that the swept-up mass from the PWN shell reaches its asymptotic value only at very late times.
As a consequence the quantity $\LztauzEFF$, deriving from Eq.~\ref{eq:Masympt}, is a function of the mass the shell has reached at the beginning of reverberation: this generally depends both on $\Lz$ and $\tauz$, and it always holds that $\LztauzEFF<\Lztauz$, while $\LztauzEFF\simeq\Lztauz$ only for $\tauz\ll1$.
Another difference is that, even if at $t\gg\tauz$ the shell expands linearly with time, at earlier times its expansion was accelerated: this makes the expansion age of the shell to be always smaller than the SNR age, even if this difference vanishes very rapidly for small $\tauz$.
The final difference, of course, is that the shell thickness is not infinitesimal.
Given all this, we have found sufficient to apply the correction for $\LztauzEFF$ to get a rather accurate approximation, at least for not too large values of $\tauz$.
For larger $\tauz$ the approximation is less accurate, but in the view that it is anyway much better than other recipes used in the past, we have decided to adopt it for all of our thin-shell models.

\begin{figure*}
\centering
	\includegraphics[width=.92\textwidth]{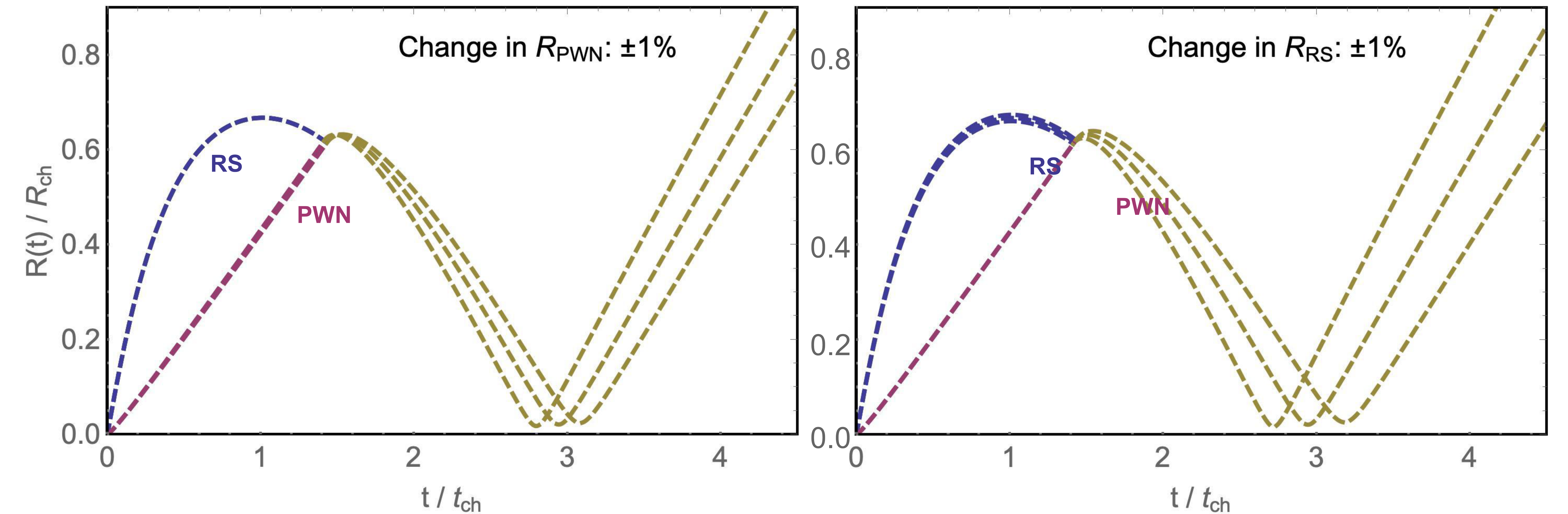}
    \caption{Variation of the PWN radius with time. The two plots illustrate how sensitive the evolution of the PWN is, during reverberation, to the profiles at $\tbegrev$. A very small variation of only $\pm 1\%$ in the RS radius (left) or in the PWN one (right) produces a strong difference in the PWN evolution later on.
    This mostly affects the time at which the minimum radius is reached, but it also changes to some extent the the PWN radius at maximum compression.}
\label{fig:dependent_on_initial}	
\end{figure*}

Fig.~\ref{fig:dependent_on_initial} gives an idea of how sensitive the PWN evolution is to the initial conditions at $\tbegrev$.
In fact even a variation of just $\pm 1\%$ in the RS or in the PWN radius at $t<\tbegrev$ reflects in a clear difference in the evolution of the PWN during reverberation.
This implies a different duration of the reverberation as well as a different level of compression, and justifies our need to improve as much as possible the analysis of the early evolution, in order to carefully get the initial conditions for the reverberation phase, and then to reliably reproduce this later phase.

\subsection{Effective mass in the shell}

The other issue, as discussed in reference to Fig.~\ref{fig:non-thinS}, is the implications of neglecting the shell thickness. The more relevant point is at the maximum of compression, because at that time the ``thin-shell'' is no longer thin, but has an width even larger than its average radial coordinate.
The finite thickness leads, at $t\sim \tbegrev$, to a slightly more efficient deceleration than if the shell mass was infinitesimally thin.
We can alternatively describe this effect by introducing an ``effective mass'' for the shell, which is slightly smaller than its actual mass: we can then find that, for small values of $\tauz$, an effective mass lower by just $\sim5\%$ than the actual mass is adequate to correctly reproduce the phase in which the original shell expansion is reverted to contraction.
Instead, in order to mimic the evolution near the maximum compression, the needed effective mass during that phase must be considerably smaller than the actual one, especially for those cases with large compression.
This is a different way to express the same concept: especially near the time of maximum compression the shell of swept-up mass may behave very different from a thin-shell.
In addition, we can see that (at least for small values of $\tauz$, where our original assumptions are better satisfied, and where a stronger compression is expected) the PWN pressure turns out to be more effective to stop the compression than predicted by the thin-shell models.
As a consequence, in general one finds that the thin-layer models tend to overestimate this compression. 
We study this in detail next.

\begin{figure*}
\centering
	\includegraphics[width=.9\textwidth]{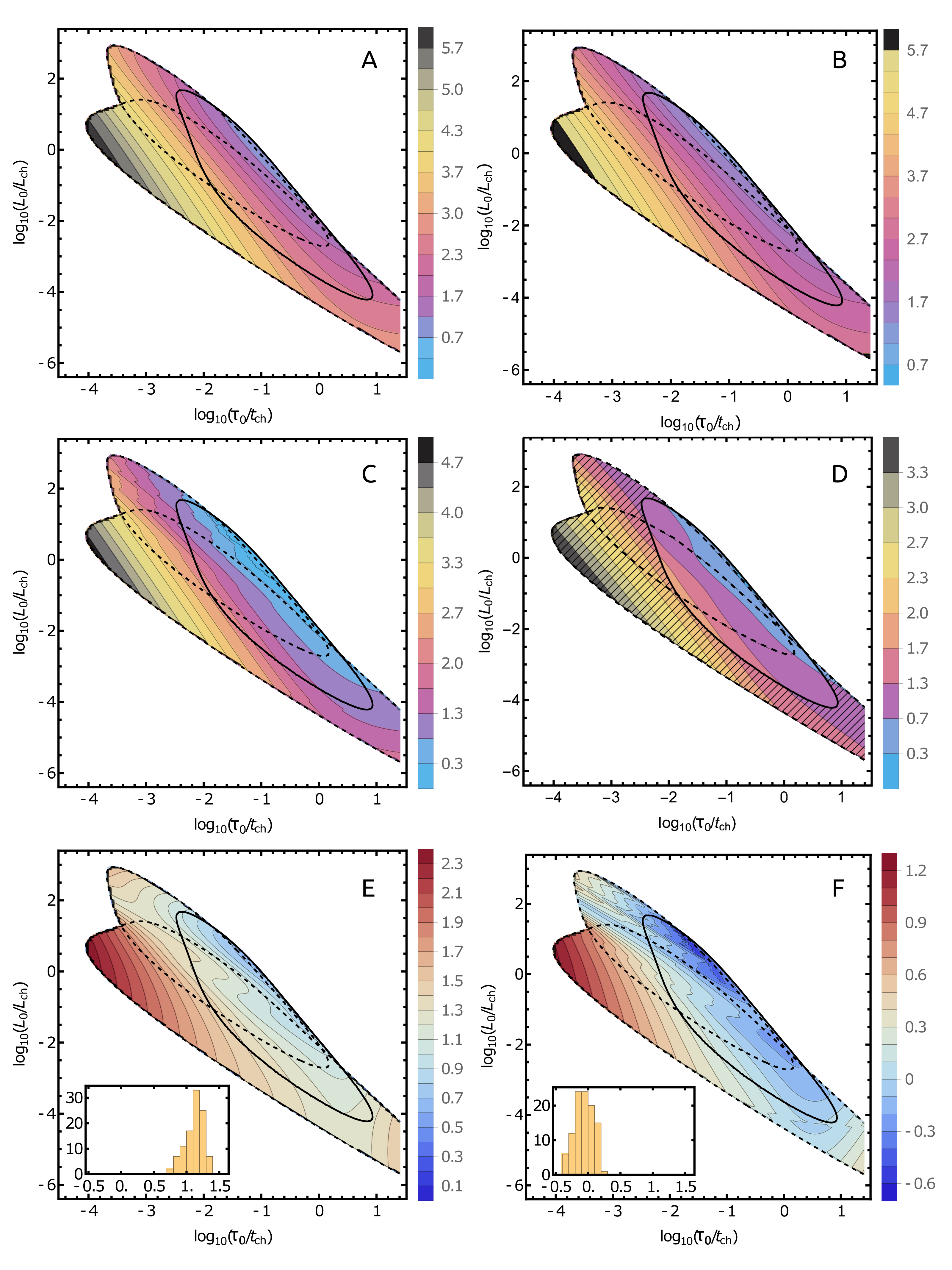}
    \caption{Maps of $\lgCF$ in the parameter space, respectively using a thin-shell model with $\Pouter=\Psedov$ (Panel A); with $\Pouter\simeq0.23\,\rho_0 V^2_{\mathrm{FS}}$ (Panel B); using our prescription for $\Pouter$ (Panel C); or with using our interpolating formula to our lagrangian models (Panel D).
    Black contours mark the position of the different populations described in Fig.~\ref{fig:moredistributions} (ours in solid). Numerical values of $\lgCF$ can be inferred in each panel by the respective vertical color bar (note that the scale is different for different plots).
    The bottom-row panels show the $\lgCF$ deviations with respect to the results of the numerical simulations, both for $\Pouter=\Psedov$ (Panel E), and for our prescription for $\Pouter$ (Panel F). The ratio of the two maps is given in logarithmic scale and also in these cases numerical values can be derived from the vertical color bar. The inset  are the histogram of the logarithmic ratios evaluated  at the points of our lagrangian models (they show a lower spread than the associated image because the former one refers to a smaller region).
    }
\label{fig:All_CF}	
\end{figure*}
\begin{figure}
\centering
	\includegraphics[width=0.46\textwidth]{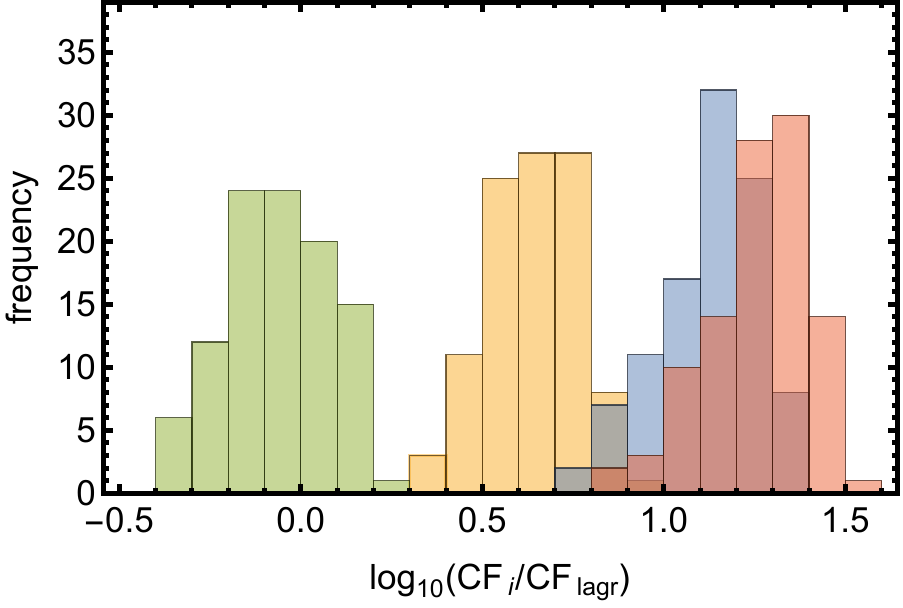}
    \caption{Histogram of the decimal logarithm of the CF ratio between thin-shell model and lagrangian models, using: our prescription for $\Pouter$ (left, in green color, which peaks around $\mathrm{CF} /\mathrm{CF}_{\mathrm{lagr}} \sim 1$, showing good agreement with the numerical models);
    the standard prescription making use of a scaling with the Sedov solution (right, in blue considering $\Pouter=\Psedov$, in red $\Pouter\simeq0.23\,\rho_0 V^2_{\mathrm{FS}}$). The intermediate case, in yellow, has been obtained considering $\Pouter=0.3\,\Psedov$ for illustrative reasons.
    }
\label{fig:CFR_H}	
\end{figure}

\section{Estimating the compression factor}
\label{sec:results}

As already mentioned, our main goal here is to estimate the magnitude of the CF, depending on the model parameters, to do it with a reasonable accuracy, and over a parameters region wide enough to cover all physically reasonable cases.
Keep in mind that no radiative losses are herein considered.
The covered area is the union of the regions gathering the 98\% of cases, for the synthesized populations from \citet{FGK:2006}, from \citet{Watters:2011}, as well as from  our choice (see Sec.~\ref{sec:pplane}).

\subsection{Variations in the compression factor due to the adopted approach}

The main results of this work are synthesized in Fig. \ref{fig:All_CF}, in which one can compare the maps of $\lgCF$ as obtained with 4 different approaches:
\begin{enumerate}[label=(\Alph*)]
    \item thin-shell with $\Pouter=\Psedov$ (i.e. the pressure derived at the center of the Sedov solution); 
    \item thin-shell with $\Pouter=0.306\,(3/4 \,\rho_0 V^2_{\mathrm{FS}})$, where $V_{\mathrm{FS}}$ is the velocity computed at the SNR FS using the trajectory given in our Paper 0; 
    \item thin-shell using our prescription for $\Pouter$ (i.e. using Eqs.~\ref{eq:aouter_first}-\ref{eq:Rshell_fit} and Appendix \ref{sec:acc-formula}); 
    \item from the lagrangian models (with extrapolation, using Eqs.~\ref{eq:CFlow} and \ref{eq:CFfull}).
\end{enumerate}
In panel D for clarity we have hatched the region outside the area sampled by numerical data.
It is apparent that using the central pressure from  the Sedov solution (panel A) the CF may reach extremely large values ($>10^5$ in the leftmost allowed region).
Rather similar results are also obtained considering the pressure computed from the trajectory of the FS (panel B).
With our prescription for the outer pressure (panel C) the CF values decrease even more, reaching large values ($\sim10^5$) at the same location of previous cases.
As a comparison the CF from numerical simulations reaches, in the same location of the parameter plane, values  up to $\sim4\E{3}$.
We believe that, in this region of extreme compression, the CF values should be somewhere in between the last two cases, possibly $\sim10^4$.

To better outline how the predictions of the various thin-shell models behave, the last two panels (E and F) show (again in logarithmic scale) the ratio between the CF from panel A and B, respectively, against the analytic approximation of the data (panel D).
From Panel E, we may see how an external $\Psedov$ always leads to overestimate the CF, only mildly for the highest energy cases, but over two orders of magnitude for the lowest energy ones.
From Panel F, instead, one may notice that over most of the allowed region the CF is approximated within about a factor 4, while only in the most extreme cases the predicted CF exceeds by an order of magnitude the extrapolation of our fit to the lagrangian models.
Both panel E and F also contain an inset with the histogram of the logarithmic ratios, evaluated at the points of our lagrangian models, and directly using the numerical data (the fact that each histogram shows a lower spread than the associated image is because the former one refers to a smaller region).
The same information is also shown Fig.~\ref{fig:CFR_H}. 
From this figure it is apparent how, on this sample, our prescription does not produce any sensible bias (average $-0.06$ and standard deviation $0.14$ in the logarithmic scale) in comparison with the numerical results, while the scalings with the Sedov solution lead to consistent overestimates (with a shift of $1.1-1.3$ in logarithmic scale).

\section{A simplified thin-shell model}
\label{sec:simplified}

A simplified thin-shell approach can be devised, in the case of small values of $\tauz$ and consequently large CF, if the only purpose is to estimate the minimum radius of the first PWN compression.
Here we exploit the idea that when the original massive shell expansion is reverted into contraction, the SNR outer pressure is by far the dominant one.
The PWN pressure plays a role in stopping the shell contraction, and reverting it again in an expansion, only when the shell is near to its final implosion, and when the outer pressure is no longer effective.
In this case we do not even need to know in detail how the SNR push evolves with time, but only what is the final implosion velocity of the shell, and this information can be derived directly from our numerical models.

It can be shown that, when a shell with an implosion velocity $\Vimplo$, interacting solely with the PWN, is decelerated by the PWN pressure, the implosion reverts to expansion at a radius:
\begin{equation}
    \label{eq:Rmindef}
    \Rmin(\Lztauz,\tauz)=\frac{2\Qbegrev(\Lztauz,\tauz)}{\Mbegrev(\Lztauz,\tauz)\Vimplo(\LztauzEFF)^2},
\end{equation}
where $\Mbegrev$ and $\Qbegrev$ are the shell mass and the quantity $Q$, evaluated at the beginning of the reverberation phase.
This formula can be easily understood by taking its similarity to the case of a mass moving, in 1D, subject to a repulsive electric field: when the mass approaches the center, its kinetic energy ($M\,V^2/2$) is converted into potential energy (in this case $Q/R$), and the minimum distance is estimated as the point at which the mass is at rest.

Once the time $\tbegrev$ is computed, by imposing $R\rs{RS}(\tbegrev)=\Rpwn(\tbegrev)$, then $\Mbegrev$ and $\Qbegrev$ can be easily derived, using the formulae we have given in Sec.~\ref{sec:pre-revPWN}.
$\Vimplo$ is a function of the effective value of $\Lztauz$, because it is derived by interpolating between our shell+SNR models; and we have also derived an approximating formula for the maximum radius, in the shell+SNR case (both formulae are given in Sec.~\ref{sec:acc-formula}, in a similar way to other quantities, discussed in Sec.~\ref{sec:shell}).
In the limit of very small $\tauz$, the quantities $\Mbegrev$ and $\Qbegrev$ can be approximated by their asymptotic values, $\Minf$ and $\Qinf$ (as from Eqs.~\ref{eq:Masympt} and \ref{eq:Qasympt}), while $\LztauzEFF$ is well approximated by $\Lztauz$.
Therefore under this limit in which $\Qinf(\Lztauz,\tauz)$ is linearly proportional to $\tauz$, the quantity $\Rmin/\tauz$ is a function of $(\Lztauz)$ only.

The method described above is an alternative, simpler way to evaluate $\Rmin$, and using an interpolation for $\Rmax$, to estimate also the CF.
The accuracy is however lower, typically about 30\%, compared with a full calculation of the thin-shell model, essentially because this simplified approach neglects further contributions to the quantity $Q(t)$ during the reverberation phase; but it is anyway interesting to better understand how $\Rmin$ is determined, and how it roughly scales with the model parameters.

\section{Conclusions}
\label{sec:end}

The evolution of a PWN+SNR system is in general a complex problem.
Due to  this, the community have adopted simplified approaches in which spherically symmetric models have been used, where the PWN is approximated by a one-zone description, the shell of swept-up material is approximated by a thin-shell model and the evolution of the pressure in the outer SNR by scaling with the Sedov solution: all of this with slightly different flavors from paper to paper.

We have tried to reduce this set of assumptions, particularly putting attention to the treatment of thin-shell features and to the evolution of the outer pressure.
Still it could be argued that spherical symmetry is not a valid assumption in probably any case, either because the explosion and/or the ambient medium are strongly anisotropic; or because even slight original anisotropies are enhanced during a strong compression phase; or because even in the case of a large scale symmetry dynamical instabilities may break the assumption of laminar flow.
It is however true that the assumption of spherical treatment is what allows one to even try of 
approaching the problem with a mixture of semi-analytical methods, and to extract valuable physical information
which we hope is general in character.

Even keeping in mind the aforementioned limitations of spherically symmetric models, in this work we have presented a thorough, and self-consistent analysis of this simplified case, giving a number of prescriptions that could be followed also in future modeling, discussing the weak aspects as well as warning against misleading approximations.
In addition, both in the main text and in the Appendices we have provided formulae that  allow one to construct a relatively simple and accurate approximation to  the PWN evolution in the pre-reverberation phase, as well as to have an approximation to the outer pressure exerted by the SNR, under a wide range of conditions.\\

\noindent To summarize, the main lessons from the present work are:\\

\noindent(i) The thin-shell approximation is very accurate to model the evolution of the PWN in the pre-reverberation phase, namely before it encounters the SNR RS shock, provided that one does not need to investigate explicitly the inner structure of the swept-up shell.\\

\noindent(ii) The situation changes considerably during the reverberation phase, because the thin-shell may be no longer a good approximation.
This is partially due to the shocked SNR material impinging on the shell during this phase that does not stick to it,  as well as to the passage of secondary shocks through the shell that may heat it up and cause it to inflate.  And finally, mostly because when the radius of the shell is compressed considerably, the ratio between the shell thickness and the nebula radius may easily be no longer small.\\

\noindent(iii) A thin-shell approximation during this phase may still be used, but one must be aware that it will tend to underestimate the shell size at the time of maximum compression, and then to  significantly overestimate the compression factor, especially in those cases subject to a strong compression.\\

\noindent(iv) The SNR pressure at the interface with the shell  must depend on the properties of the shell itself (which can be identified with its mass).
The idea of a SNR pressure independent of its interaction with the shell is conceptually wrong.\\

\noindent(v) The usage of a Sedov model for scaling the outer pressure, not only is unjustified, but may lead to very unreliable results.  We have shown that one cannot use a pressure derived from the Sedov solution, since it is valid only at much later times, when most of the SNR mass consists of ambient medium, rather than of ejecta.
The Sedov model does not apply to a SNR when it still is essentially made of stellar ejecta; moreover the SNR here is interacting with the PWN, and therefore the structure of its pressure will change depending on how strong this interaction is.\\

\noindent(vi) A still approximated case, but much more accurate than the Sedov assumption, is that derived from numerical solutions of the interaction of a massive expanding shell with a SNR. These models form a one-parameter family of solutions, which depends only on the mass of the shell (in units of the mass of the SNR ejecta).
This approximation is well motivated for all cases with $\tauz$ smaller than the characteristic time of the SNR evolution ($\tch$), because most of the information about the early PWN evolution is condensed at later times in the shell mass.
A comparison between the  pressure evolution for some of these models and the Sedov case is sufficient to get an idea of how bad the Sedov assumption may be.\\

\noindent(vii) To further complicate the modeling of the PWN evolution during reverberation is the fact that it is very sensitive to various details: not only the value of the outer pressure, but also the pre-reverberation evolution.\\

\noindent(viii) Apart for the cases with large $\tauz$ values, of less importance here because they do not involve a high compression of the PWN, in all other cases the massive shell has the role of a ``buffer'', an interface between the PWN and SNR dynamical pressure forces: first the PWN pressure accelerate the shell; then the SNR pressure acts to revert the expansion of the shell size into compression, and during this phase the PWN dynamical effect is negligible; after then the SNR pressure will become unimportant, and the dynamics of the shell at the time of its maximum compression will be ruled by the interaction between the PWN pressure and the inertia of the converging shell.
\\

\noindent The present work is part of a more extended project.
The aim is to understand nebula reverberation, as it is critical not only to be able to analyze in depth the PWN/SNR/ISM connection but for evolving these systems beyond a few thousand years, when they are mostly observed with instrumentation at different frequencies. 
One of our earlier works \citep{Bandiera:2021} dealt with providing highly accurate approximations for the evolution of the main structural features of supernova remnants, such as RS, FS, and CD. We compared our results with previously adopted approximations, showing that existing simplified prescriptions can easily lead to large errors.
We also provided highly accurate approximations for the initial phase of evolution as well. 
In \citet{Bandiera:2020}  we investigated how different prescriptions for various parameters, like the properties of the supernova ejecta, of the inner pulsar, as well of the ambient medium, shift the starting time of the reverberation phase, how this affects the amount of the compression, and how much of this can be ascribable to the radiation processes. 
We still kept several of the approximations we dealt with here,
but it was then clear that it was crucial to find a good representation of the pressure of the ejecta, remarking that after the RS and the PWN collide, radiative models assuming their existence as separate entities are necessarily inaccurate. We realized in particular that the assumption of the bounding SNR to be in the relaxed Sedov state must be handled with care, since the dynamics of the swept-up shell appears in fact to be very sensitive to the ejecta profiles.
This is what we intended to cover in this work.

We expect that the next steps will consist in joining our know-how about numerical and semi-analytic models for computing models that will account for the inner structure of the swept-up shell and at the same time will improve the initial conditions, in order to properly treat also cases with small $\tauz$, and finally adding the dynamical effects of the (synchrotron) radiation losses inside the PWN.
Radiation losses near the time of maximum compression are expected to behave like a sort of threshold effect.
For this reason, a factor of a few in the compression factor, when radiation effects are not included, may lead to very different behaviours when radiation is included in the modeling. When radiation starts to be effective, it may lead to a sort of avalanche effect, by decreasing the PWN pressure, therefore enhancing compression and increasing the magnetic strength, and in turn further increasing the radiation losses.

Despite essentially all of the intricacies of the dynamical aspects of the reverberation phase we studied here are simply ignored in the literature
when reverberation is acknowledged --not the common case either--, 
it is usual to find modeling and predictions (from single PWNe to population studies) for PWNe of tens of thousand of years.
As a final conclusion, then, we state that in our opinion, without having a consistent model for the reverberation phase --properly including all dynamical aspects together with radiation-- predictions for middle-age PWNe beyond reverberation are most likely meaningless. 
On the contrary, proper PWNe radiation modeling has yet to pass few thousands years.
Our future works will be devoted to that.

\section*{Acknowledgements}
%
This work has been supported by INAF grants MAINSTREAM 2018, SKA-CTA, PRIN-INAF 2019, by ASI-INAF n.2017-14-H.O, grant PID2021-124581OB-I00 and by the Spanish program Unidad de Excelencia María de Maeztu CEX2020-001058-M.

\section*{Data availability}
The data for the models underlying this article will be shared on reasonable request to the corresponding author.


\bibliographystyle{mn2e} 
\bibliography{biblio}

\begin{thebibliography}{}

\bibitem[\protect\citeauthoryear{{Aharonian}, {An}, {Axikegu}, {Bai}, {Bai} \&
  {LHAASO Collaboration}}{{Aharonian} et~al.}{2020}]{Aharonian2020}
{Aharonian} F.,  {An} Q.,  {Axikegu} {Bai} L.~X.,  {Bai} Y.~X.,    {LHAASO
  Collaboration} 2020, arXiv e-prints, p. arXiv:2010.06205

\bibitem[\protect\citeauthoryear{{Amato} \& {Blasi}}{{Amato} \&
  {Blasi}}{2018}]{Amato:2018}
{Amato} E.,  {Blasi} P.,  2018, Advances in Space Research, 62, 2731

\bibitem[\protect\citeauthoryear{{Arons}}{{Arons}}{2012}]{Arons:2012}
{Arons} J.,  2012, \ssr, 173, 341

\bibitem[\protect\citeauthoryear{{Asvarov}}{{Asvarov}}{2014}]{Asvarov:2014}
{Asvarov} A.~I.,  2014, \aap, 561, A70

\bibitem[\protect\citeauthoryear{{Badenes}, {Maoz} \& {Draine}}{{Badenes}
  et~al.}{2010}]{Badenes:2010}
{Badenes} C.,  {Maoz} D.,    {Draine} B.~T.,  2010, \mnras, 407, 1301

\bibitem[\protect\citeauthoryear{{Bandiera}, {Bucciantini}, {Mart{\'\i}n},
  {Olmi} \& {Torres}}{{Bandiera} et~al.}{2020}]{Bandiera:2020}
{Bandiera} R.,  {Bucciantini} N.,  {Mart{\'\i}n} J.,  {Olmi} B.,    {Torres}
  D.~F.,  2020, \mnras, 499, 2051

\bibitem[\protect\citeauthoryear{{Bandiera}, {Bucciantini}, {Mart{\'\i}n},
  {Olmi} \& {Torres}}{{Bandiera} et~al.}{2021}]{Bandiera:2021}
{Bandiera} R.,  {Bucciantini} N.,  {Mart{\'\i}n} J.,  {Olmi} B.,    {Torres}
  D.~F.,  2021, \mnras, 508, 3194

\bibitem[\protect\citeauthoryear{{Bandiera} \& {Petruk}}{{Bandiera} \&
  {Petruk}}{2010}]{Bandiera_Petruk:2010}
{Bandiera} R.,  {Petruk} O.,  2010, \aap, 509, A34

\bibitem[\protect\citeauthoryear{{Berkhuijsen}}{{Berkhuijsen}}{1987}]{Berkhuijsen:1987}
{Berkhuijsen} E.~M.,  1987, \aap, 181, 398

\bibitem[\protect\citeauthoryear{{Blasi} \& {Amato}}{{Blasi} \&
  {Amato}}{2011}]{Blasi:2011}
{Blasi} P.,  {Amato} E.,  2011, ArXiv:1007.4745

\bibitem[\protect\citeauthoryear{{Blondin}, {Chevalier} \&
  {Frierson}}{{Blondin} et~al.}{2001}]{Blondin:2001}
{Blondin} J.~M.,  {Chevalier} R.~A.,    {Frierson} D.~M.,  2001, \apj, 563, 806

\bibitem[\protect\citeauthoryear{{Bucciantini}, {Amato}, {Bandiera}, {Blondin}
  \& {Del Zanna}}{{Bucciantini} et~al.}{2004}]{Bucciantini:2004a}
{Bucciantini} N.,  {Amato} E.,  {Bandiera} R.,  {Blondin} J.~M.,    {Del Zanna}
  L.,  2004, \aap, 423, 253

\bibitem[\protect\citeauthoryear{{Bucciantini}, {Arons} \&
  {Amato}}{{Bucciantini} et~al.}{2011}]{Bucciantini_Arons+11a}
{Bucciantini} N.,  {Arons} J.,    {Amato} E.,  2011, \mnras, 410, 381

\bibitem[\protect\citeauthoryear{{Bucciantini}, {Bandiera}, {Blondin}, {Amato}
  \& {Del Zanna}}{{Bucciantini} et~al.}{2004}]{Bucciantini:2004b}
{Bucciantini} N.,  {Bandiera} R.,  {Blondin} J.~M.,  {Amato} E.,    {Del Zanna}
  L.,  2004, \aap, 422, 609

\bibitem[\protect\citeauthoryear{{Bucciantini}, {Blondin}, {Del Zanna} \&
  {Amato}}{{Bucciantini} et~al.}{2003}]{Bucciantini:2003}
{Bucciantini} N.,  {Blondin} J.~M.,  {Del Zanna} L.,    {Amato} E.,  2003,
  \aap, 405, 617

\bibitem[\protect\citeauthoryear{{Bucciantini}, {Thompson}, {Arons}, {Quataert}
  \& {Del Zanna}}{{Bucciantini} et~al.}{2007}]{Bucciantini_Thompson+07a}
{Bucciantini} N.,  {Thompson} T.~A.,  {Arons} J.,  {Quataert} E.,    {Del
  Zanna} L.,  2007, Advances in Space Research, 40, 1441

\bibitem[\protect\citeauthoryear{{Chevalier}}{{Chevalier}}{1982}]{Chevalier1982}
{Chevalier} R.~A.,  1982, \apj, 258, 790

\bibitem[\protect\citeauthoryear{{Chevalier}}{{Chevalier}}{2005}]{Chevalier2005}
{Chevalier} R.~A.,  2005, \apj, 619, 839

\bibitem[\protect\citeauthoryear{{Chevalier} \& {Soker}}{{Chevalier} \&
  {Soker}}{1989}]{Chevalier:1989}
{Chevalier} R.~A.,  {Soker} N.,  1989, \apj, 341, 867

\bibitem[\protect\citeauthoryear{{Contopoulos}, {Kazanas} \&
  {Fendt}}{{Contopoulos} et~al.}{1999}]{Contopoulos_Kazanas+99a}
{Contopoulos} I.,  {Kazanas} D.,    {Fendt} C.,  1999, \apj, 511, 351

\bibitem[\protect\citeauthoryear{{de O{\~n}a-Wilhelmi}, {Rudak}, {Barrio},
  {Contreras}, {Gallant}, {Hadasch}, {Hassan}, {Lopez}, {Mazin}, {Mirabal},
  {Pedaletti}, {Renaud}, {de los Reyes}, {Torres} \& {CTA Consortium}}{{de
  O{\~n}a-Wilhelmi} et~al.}{2013}]{de-Ona-Wilhelmi:2013}
{de O{\~n}a-Wilhelmi} E.,  {Rudak} B.,  {Barrio} J.~A.,  {Contreras} J.~L.,
  {Gallant} Y.,  {Hadasch} D.,  {Hassan} T.,  {Lopez} M.,  {Mazin} D.,
  {Mirabal} N.,  {Pedaletti} G.,  {Renaud} M.,  {de los Reyes} R.,  {Torres}
  D.~F.,    {CTA Consortium} 2013, Astroparticle Physics, 43, 287

\bibitem[\protect\citeauthoryear{{Faucher-Gigu{\`e}re} \&
  {Kaspi}}{{Faucher-Gigu{\`e}re} \& {Kaspi}}{2006}]{FGK:2006}
{Faucher-Gigu{\`e}re} C.-A.,  {Kaspi} V.~M.,  2006, \apj, 643, 332

\bibitem[\protect\citeauthoryear{{Fiori}, {Olmi}, {Amato}, {Bandiera},
  {Bucciantini}, N. {Zampieri} \& {Burtovoi}}{{Fiori}
  et~al.}{2021}]{Fiori:2021}
{Fiori} M.,  {Olmi} B.,  {Amato} E.,  {Bandiera} R.,  {Bucciantini} N.
  {Zampieri} L.,    {Burtovoi} A.,  2021, To be sumbitted to \mnras

\bibitem[\protect\citeauthoryear{{Gaensler} \& {Slane}}{{Gaensler} \&
  {Slane}}{2006}]{Gaensler_Slane06a}
{Gaensler} B.~M.,  {Slane} P.~O.,  2006, \araa, 44, 17

\bibitem[\protect\citeauthoryear{{Gelfand}, {Slane} \& {Zhang}}{{Gelfand}
  et~al.}{2009}]{Gelfand_Slane+09a}
{Gelfand} J.~D.,  {Slane} P.~O.,    {Zhang} W.,  2009, \apj, 703, 2051

\bibitem[\protect\citeauthoryear{Guderley}{Guderley}{1942}]{Guderley:1942}
Guderley K.~G.,  1942, Luftfahrtforschung, 19, 302

\bibitem[\protect\citeauthoryear{{Gull{\'o}n}, {Miralles}, {Vigan{\`o}} \&
  {Pons}}{{Gull{\'o}n} et~al.}{2014}]{Gullon:2014}
{Gull{\'o}n} M.,  {Miralles} J.~A.,  {Vigan{\`o}} D.,    {Pons} J.~A.,  2014,
  \mnras, 443, 1891

\bibitem[\protect\citeauthoryear{{H.~E.~S.~S.~Collaboration}, {Abdalla},
  {Abramowski}, {Aharonian}, {Ait Benkhali}, {Akhperjanian}, {Andersson},
  {Ang{\"u}ner}, {Arrieta}, {Aubert} \& et al.}{{H.~E.~S.~S.~Collaboration}
  et~al.}{2018}]{HESS-PWN-2018}
{H.~E.~S.~S.~Collaboration} {Abdalla} H.,  {Abramowski} A.,  {Aharonian} F.,
  {Ait Benkhali} F.,  {Akhperjanian} A.~G.,  {Andersson} T.,  {Ang{\"u}ner}
  E.~O.,  {Arrieta} M.,  {Aubert} P.,    et al. 2018, \aap, 612, A2

\bibitem[\protect\citeauthoryear{{Hamilton} \& {Sarazin}}{{Hamilton} \&
  {Sarazin}}{1984}]{Hamilton_Sarazin:1984}
{Hamilton} A.~J.~S.,  {Sarazin} C.~L.,  1984, \apj, 281, 682

\bibitem[\protect\citeauthoryear{{Hester}}{{Hester}}{2008}]{Hester:2008}
{Hester} J.~J.,  2008, \araa, 46, 127

\bibitem[\protect\citeauthoryear{{Horvath}}{{Horvath}}{2019}]{Horvath:2019}
{Horvath} J.~E.,  2019, \mnras, 484, 1983

\bibitem[\protect\citeauthoryear{{Johnston}, {Smith}, {Karastergiou} \&
  {Kramer}}{{Johnston} et~al.}{2020}]{Johnston:2020}
{Johnston} S.,  {Smith} D.~A.,  {Karastergiou} A.,    {Kramer} M.,  2020,
  \mnras

\bibitem[\protect\citeauthoryear{{Jun}}{{Jun}}{1998}]{Jun1998}
{Jun} B.-I.,  1998, \apj, 499, 282

\bibitem[\protect\citeauthoryear{{Karamehmetoglu}, {Taddia}, {Sollerman},
  {Wyrzykowski}, {Schmidl}, {Fraser}, {Fremling}, {Greiner}, {Inserra},
  {Kostrzewa-Rutkowska}, {Maguire}, {Smartt}, {Sullivan} \&
  {Young}}{{Karamehmetoglu} et~al.}{2017}]{Karamehmetoglu:2017}
{Karamehmetoglu} E.,  {Taddia} F.,  {Sollerman} J.,  {Wyrzykowski} {\L}.,
  {Schmidl} S.,  {Fraser} M.,  {Fremling} C.,  {Greiner} J.,  {Inserra} C.,
  {Kostrzewa-Rutkowska} Z.,  {Maguire} K.,  {Smartt} S.,  {Sullivan} M.,
  {Young} D.~R.,  2017, \aap, 602, A93

\bibitem[\protect\citeauthoryear{{Kargaltsev} \& {Pavlov}}{{Kargaltsev} \&
  {Pavlov}}{2008}]{Kargaltsev2008}
{Kargaltsev} O.,  {Pavlov} G.~G.,  2008, in {Bassa} C.,  {Wang} Z.,  {Cumming}
  A.,   {Kaspi} V.~M.,  eds, 40 Years of Pulsars: Millisecond Pulsars,
  Magnetars and More Vol.~983 of American Institute of Physics Conference
  Series, {Pulsar Wind Nebulae in the Chandra Era}.
pp 171--185

\bibitem[\protect\citeauthoryear{{Kargaltsev}, {Pavlov}, {Klingler} \&
  {Rangelov}}{{Kargaltsev} et~al.}{2017}]{Kargaltsev2017}
{Kargaltsev} O.,  {Pavlov} G.~G.,  {Klingler} N.,    {Rangelov} B.,  2017,
  Journal of Plasma Physics, 83, 635830501

\bibitem[\protect\citeauthoryear{{Kasen}}{{Kasen}}{2010}]{Kasen:2010}
{Kasen} D.,  2010, \apj, 708, 1025

\bibitem[\protect\citeauthoryear{{Kennel} \& {Coroniti}}{{Kennel} \&
  {Coroniti}}{1984}]{Kennel_Coroniti84b}
{Kennel} C.~F.,  {Coroniti} F.~V.,  1984, \apj, 283, 710

\bibitem[\protect\citeauthoryear{{Klepser}, {Carrigan}, {de O{\~n}a Wilhelmi},
  {Deil}, {F{\"o}rster}, {Marandon}, {Mayer}, {Stycz}, {Valerius} \&
  {H.~E.~S.~S. Collaboration}}{{Klepser} et~al.}{2013}]{Klepser:2013}
{Klepser} S.,  {Carrigan} S.,  {de O{\~n}a Wilhelmi} E.,  {Deil} C.,
  {F{\"o}rster} A.,  {Marandon} V.,  {Mayer} M.,  {Stycz} K.,  {Valerius} K.,
   {H.~E.~S.~S. Collaboration} 2013, in International Cosmic Ray Conference
  Vol.~33 of International Cosmic Ray Conference, {A Population of
  Teraelectronvolt Pulsar Wind Nebulae in the H.E.S.S. Galactic Plane Survey}.
p.~971

\bibitem[\protect\citeauthoryear{{Kolb}, {Blondin}, {Slane} \& {Temim}}{{Kolb}
  et~al.}{2017}]{Kolb:2017}
{Kolb} C.,  {Blondin} J.,  {Slane} P.,    {Temim} T.,  2017, \apj, 844, 1

\bibitem[\protect\citeauthoryear{{Komissarov}}{{Komissarov}}{2001}]{Komissarov:2001}
{Komissarov} S.~S.,  2001, \mnras, 326, L41

\bibitem[\protect\citeauthoryear{{Kurf{\"u}rst}, {Pejcha} \&
  {Krti{\v{c}}ka}}{{Kurf{\"u}rst} et~al.}{2020}]{Kurfurst:2020}
{Kurf{\"u}rst} P.,  {Pejcha} O.,    {Krti{\v{c}}ka} J.,  2020, \aap, 642, A214

\bibitem[\protect\citeauthoryear{{Long}, {Blair}, {Winkler}, {Becker}, {Gaetz},
  {Ghavamian}, {Helfand}, {Hughes}, {Kirshner}, {Kuntz}, {McNeil}, {Pannuti},
  {Plucinsky}, {Saul}, {T{\"u}llmann} \& {Williams}}{{Long}
  et~al.}{2010}]{Long:2010}
{Long} K.~S.,  {Blair} W.~P.,  {Winkler} P.~F.,  {Becker} R.~H.,  {Gaetz}
  T.~J.,  {Ghavamian} P.,  {Helfand} D.~J.,  {Hughes} J.~P.,  {Kirshner} R.~P.,
   {Kuntz} K.~D.,  {McNeil} E.~K.,  {Pannuti} T.~G.,  {Plucinsky} P.~P.,
  {Saul} D.,  {T{\"u}llmann} R.,    {Williams} B.,  2010, \apjs, 187, 495

\bibitem[\protect\citeauthoryear{{Lyne}, {Jordan}, {Graham-Smith}, {Espinoza},
  {Stappers} \& {Weltevrede}}{{Lyne} et~al.}{2015}]{Lyne:2015}
{Lyne} A.~G.,  {Jordan} C.~A.,  {Graham-Smith} F.,  {Espinoza} C.~M.,
  {Stappers} B.~W.,    {Weltevrede} P.,  2015, \mnras, 446, 857

\bibitem[\protect\citeauthoryear{{Ma}, {Ng}, {Bucciantini}, {Slane}, {Gaensler}
  \& {Temim}}{{Ma} et~al.}{2016}]{Ma_Ng+16a}
{Ma} Y.~K.,  {Ng} C.-Y.,  {Bucciantini} N.,  {Slane} P.~O.,  {Gaensler} B.~M.,
    {Temim} T.,  2016, \apj, 820, 100

\bibitem[\protect\citeauthoryear{{Magnier}, {Primini}, {Prins}, {van Paradijs}
  \& {Lewin}}{{Magnier} et~al.}{1997}]{Magnier:1997}
{Magnier} E.~A.,  {Primini} F.~A.,  {Prins} S.,  {van Paradijs} J.,    {Lewin}
  W. H.~G.,  1997, \apj, 490, 649

\bibitem[\protect\citeauthoryear{{Manchester} \& {Peterson}}{{Manchester} \&
  {Peterson}}{1989}]{Manchester:1989}
{Manchester} R.~N.,  {Peterson} B.~A.,  1989, \apjl, 342, L23

\bibitem[\protect\citeauthoryear{{Martin} \& {Torres}}{{Martin} \&
  {Torres}}{2022}]{Martin2022j}
{Martin} J.,  {Torres} D.~F.,  2022, Journal of High Energy Astrophysics, 36,
  128

\bibitem[\protect\citeauthoryear{{Mart{\'{\i}}n}, {Torres} \&
  {Pedaletti}}{{Mart{\'{\i}}n} et~al.}{2016}]{Martin:2016}
{Mart{\'{\i}}n} J.,  {Torres} D.~F.,    {Pedaletti} G.,  2016, \mnras, 459,
  3868

\bibitem[\protect\citeauthoryear{{Mart{\'{\i}}n}, {Torres} \&
  {Rea}}{{Mart{\'{\i}}n} et~al.}{2012}]{Martin_Torres+12a}
{Mart{\'{\i}}n} J.,  {Torres} D.~F.,    {Rea} N.,  2012, \mnras, 427, 415

\bibitem[\protect\citeauthoryear{{Matzner} \& {McKee}}{{Matzner} \&
  {McKee}}{1999}]{Matzner:1999}
{Matzner} C.~D.,  {McKee} C.~F.,  1999, \apj, 510, 379

\bibitem[\protect\citeauthoryear{{Mestre}, {Torres}, {de O{\~n}a Wilhelmi} \&
  {Mart{\'\i}}}{{Mestre} et~al.}{2022}]{Mestre2022}
{Mestre} E.,  {Torres} D.~F.,  {de O{\~n}a Wilhelmi} E.,    {Mart{\'\i}} J.,
  2022, \mnras

\bibitem[\protect\citeauthoryear{{Meyer}, {Petrov} \& {Pohl}}{{Meyer}
  et~al.}{2020}]{Meyer:2020}
{Meyer} D.~M.~A.,  {Petrov} M.,    {Pohl} M.,  2020, \mnras, 493, 3548

\bibitem[\protect\citeauthoryear{{Meyer}, {Pohl}, {Petrov} \&
  {Oskinova}}{{Meyer} et~al.}{2021}]{Meyer2021}
{Meyer} D.~M.~A.,  {Pohl} M.,  {Petrov} M.,    {Oskinova} L.,  2021, \mnras,
  502, 5340

\bibitem[\protect\citeauthoryear{{Miceli}, {Orlando}, {Reale}, {Bocchino} \&
  {Peres}}{{Miceli} et~al.}{2013}]{Miceli:2013}
{Miceli} M.,  {Orlando} S.,  {Reale} F.,  {Bocchino} F.,    {Peres} G.,  2013,
  \mnras, 430, 2864

\bibitem[\protect\citeauthoryear{{Mignone}, {Bodo}, {Massaglia}, {Matsakos},
  {Tesileanu}, {Zanni} \& {Ferrari}}{{Mignone} et~al.}{2007}]{Mignone2007}
{Mignone} A.,  {Bodo} G.,  {Massaglia} S.,  {Matsakos} T.,  {Tesileanu} O.,
  {Zanni} C.,    {Ferrari} A.,  2007, \apjs, 170, 228

\bibitem[\protect\citeauthoryear{{Olmi} \& {Bucciantini}}{{Olmi} \&
  {Bucciantini}}{2019}]{Olmi_Bucciantini:2019}
{Olmi} B.,  {Bucciantini} N.,  2019, \mnras, 484, 5755

\bibitem[\protect\citeauthoryear{{Olmi}, {Del Zanna}, {Amato}, {Bucciantini} \&
  {Mignone}}{{Olmi} et~al.}{2016}]{Olmi:2016}
{Olmi} B.,  {Del Zanna} L.,  {Amato} E.,  {Bucciantini} N.,    {Mignone} A.,
  2016, Journal of Plasma Physics, 82, 635820601

\bibitem[\protect\citeauthoryear{{Olmi} \& {Torres}}{{Olmi} \&
  {Torres}}{2020}]{Olmi_Torres:2020}
{Olmi} B.,  {Torres} D.~F.,  2020, \mnras, 494, 4357

\bibitem[\protect\citeauthoryear{{Parthasarathy}, {Johnston}, {Shannon},
  {Lentati}, {Bailes}, {Dai}, {Kerr}, {Manchester}, {Os{\l}owski}, {Sobey},
  {van Straten} \& {Weltevrede}}{{Parthasarathy}
  et~al.}{2020}]{Parthasarathy:2020}
{Parthasarathy} A.,  {Johnston} S.,  {Shannon} R.~M.,  {Lentati} L.,  {Bailes}
  M.,  {Dai} S.,  {Kerr} M.,  {Manchester} R.~N.,  {Os{\l}owski} S.,  {Sobey}
  C.,  {van Straten} W.,    {Weltevrede} P.,  2020, \mnras, 494, 2012

\bibitem[\protect\citeauthoryear{{Porth}, {Komissarov} \& {Keppens}}{{Porth}
  et~al.}{2014a}]{Porth:2014a}
{Porth} O.,  {Komissarov} S.~S.,    {Keppens} R.,  2014a, \mnras, 443, 547

\bibitem[\protect\citeauthoryear{{Porth}, {Komissarov} \& {Keppens}}{{Porth}
  et~al.}{2014b}]{Porth:2014}
{Porth} O.,  {Komissarov} S.~S.,    {Keppens} R.,  2014b, \mnras, 438, 278

\bibitem[\protect\citeauthoryear{{Potter}, {Staveley-Smith}, {Reville}, {Ng},
  {Bicknell}, {Sutherland} \& {Wagner}}{{Potter} et~al.}{2014}]{Potter2014}
{Potter} T.~M.,  {Staveley-Smith} L.,  {Reville} B.,  {Ng} C.~Y.,  {Bicknell}
  G.~V.,  {Sutherland} R.~S.,    {Wagner} A.~Y.,  2014, \apj, 794, 174

\bibitem[\protect\citeauthoryear{{Rees} \& {Gunn}}{{Rees} \&
  {Gunn}}{1974}]{Rees:1974}
{Rees} M.~J.,  {Gunn} J.~E.,  1974, \mnras, 167, 1

\bibitem[\protect\citeauthoryear{{Reynolds} \& {Chevalier}}{{Reynolds} \&
  {Chevalier}}{1984}]{Reynolds:1984}
{Reynolds} S.~P.,  {Chevalier} R.~A.,  1984, \apj, 278, 630

\bibitem[\protect\citeauthoryear{{Sedov}}{{Sedov}}{1946}]{Sedov1946}
{Sedov} L.~I.,  1946, Journal of Applied Mathematics and Mechanics, 10, 241

\bibitem[\protect\citeauthoryear{{Sironi} \& {Cerutti}}{{Sironi} \&
  {Cerutti}}{2017}]{Sironi2017}
{Sironi} L.,  {Cerutti} B.,  2017, in {Torres} D.~F.,  ed., Modelling Pulsar
  Wind Nebulae Vol.~446 of Astrophysics and Space Science Library, {Particle
  Acceleration in Pulsar Wind Nebulae: PIC Modelling}.
p.~247

\bibitem[\protect\citeauthoryear{{Sironi} \& {Spitkovsky}}{{Sironi} \&
  {Spitkovsky}}{2009}]{Sironi:2009}
{Sironi} L.,  {Spitkovsky} A.,  2009, \apjl, 707, L92

\bibitem[\protect\citeauthoryear{{Smartt}}{{Smartt}}{2009}]{Smartt:2009}
{Smartt} S.~J.,  2009, \araa, 47, 63

\bibitem[\protect\citeauthoryear{{Spitkovsky}}{{Spitkovsky}}{2006}]{Spitkovsky:2006}
{Spitkovsky} A.,  2006, \apjl, 648, L51

\bibitem[\protect\citeauthoryear{{Torres}}{{Torres}}{2017}]{Torres:2017}
{Torres} D.~F.,  2017, \apj, 835, 54

\bibitem[\protect\citeauthoryear{{Torres}, {Cillis}, {Mart{\'\i}n} \& {de
  O{\~n}a Wilhelmi}}{{Torres} et~al.}{2014}]{Torres:2014}
{Torres} D.~F.,  {Cillis} A.,  {Mart{\'\i}n} J.,    {de O{\~n}a Wilhelmi} E.,
  2014, Journal of High Energy Astrophysics, 1, 31

\bibitem[\protect\citeauthoryear{{Torres} \& {Lin}}{{Torres} \&
  {Lin}}{2018}]{Torres2018b}
{Torres} D.~F.,  {Lin} T.,  2018, \apjl, 864, L2

\bibitem[\protect\citeauthoryear{{Torres}, {Lin} \& {Coti Zelati}}{{Torres}
  et~al.}{2019}]{Torres2019}
{Torres} D.~F.,  {Lin} T.,    {Coti Zelati} F.,  2019, \mnras, 486, 1019

\bibitem[\protect\citeauthoryear{{Torres}, {Mart{\'\i}n}, {de O{\~n}a Wilhelmi}
  \& {Cillis}}{{Torres} et~al.}{2013}]{Torres2013}
{Torres} D.~F.,  {Mart{\'\i}n} J.,  {de O{\~n}a Wilhelmi} E.,    {Cillis} A.,
  2013, \mnras, 436, 3112

\bibitem[\protect\citeauthoryear{{Truelove} \& {McKee}}{{Truelove} \&
  {McKee}}{1999}]{Truelove1999}
{Truelove} J.~K.,  {McKee} C.~F.,  1999, \apjs, 120, 299

\bibitem[\protect\citeauthoryear{{van der Swaluw}, {Achterberg}, {Gallant},
  {Downes} \& {Keppens}}{{van der Swaluw} et~al.}{2003}]{van-der-Swaluw:2003}
{van der Swaluw} E.,  {Achterberg} A.,  {Gallant} Y.~A.,  {Downes} T.~P.,
  {Keppens} R.,  2003, \aap, 397, 913

\bibitem[\protect\citeauthoryear{{van der Swaluw}, {Achterberg}, {Gallant} \&
  {T{\'o}th}}{{van der Swaluw} et~al.}{2001}]{van-der-Swaluw:2001}
{van der Swaluw} E.,  {Achterberg} A.,  {Gallant} Y.~A.,    {T{\'o}th} G.,
  2001, \aap, 380, 309

\bibitem[\protect\citeauthoryear{{Vorster}, {Tibolla}, {Ferreira} \&
  {Kaufmann}}{{Vorster} et~al.}{2013}]{Vorster:2013}
{Vorster} M.~J.,  {Tibolla} O.,  {Ferreira} S.~E.~S.,    {Kaufmann} S.,  2013,
  \apj, 773, 139

\bibitem[\protect\citeauthoryear{{Watters} \& {Romani}}{{Watters} \&
  {Romani}}{2011}]{Watters:2011}
{Watters} K.~P.,  {Romani} R.~W.,  2011, \apj, 727, 123

\bibitem[\protect\citeauthoryear{{Weisskopf}, {Silver}, {Kestenbaum}, {Long} \&
  {Novick}}{{Weisskopf} et~al.}{1978}]{Weisskopf1978}
{Weisskopf} M.~C.,  {Silver} E.~H.,  {Kestenbaum} H.~L.,  {Long} K.~S.,
  {Novick} R.,  1978, \apjl, 220, L117

\bibitem[\protect\citeauthoryear{{Westfold}}{{Westfold}}{1959}]{Westfold59a}
{Westfold} K.~C.,  1959, \apj, 130, 241

\end{thebibliography}

\newpage 

\appendix


\section{Time evolution of pressure and density profiles for some selected cases.}
\label{sec:app-varprofiles}
\begin{figure*}
\centering
	\includegraphics[width=0.99\textwidth]{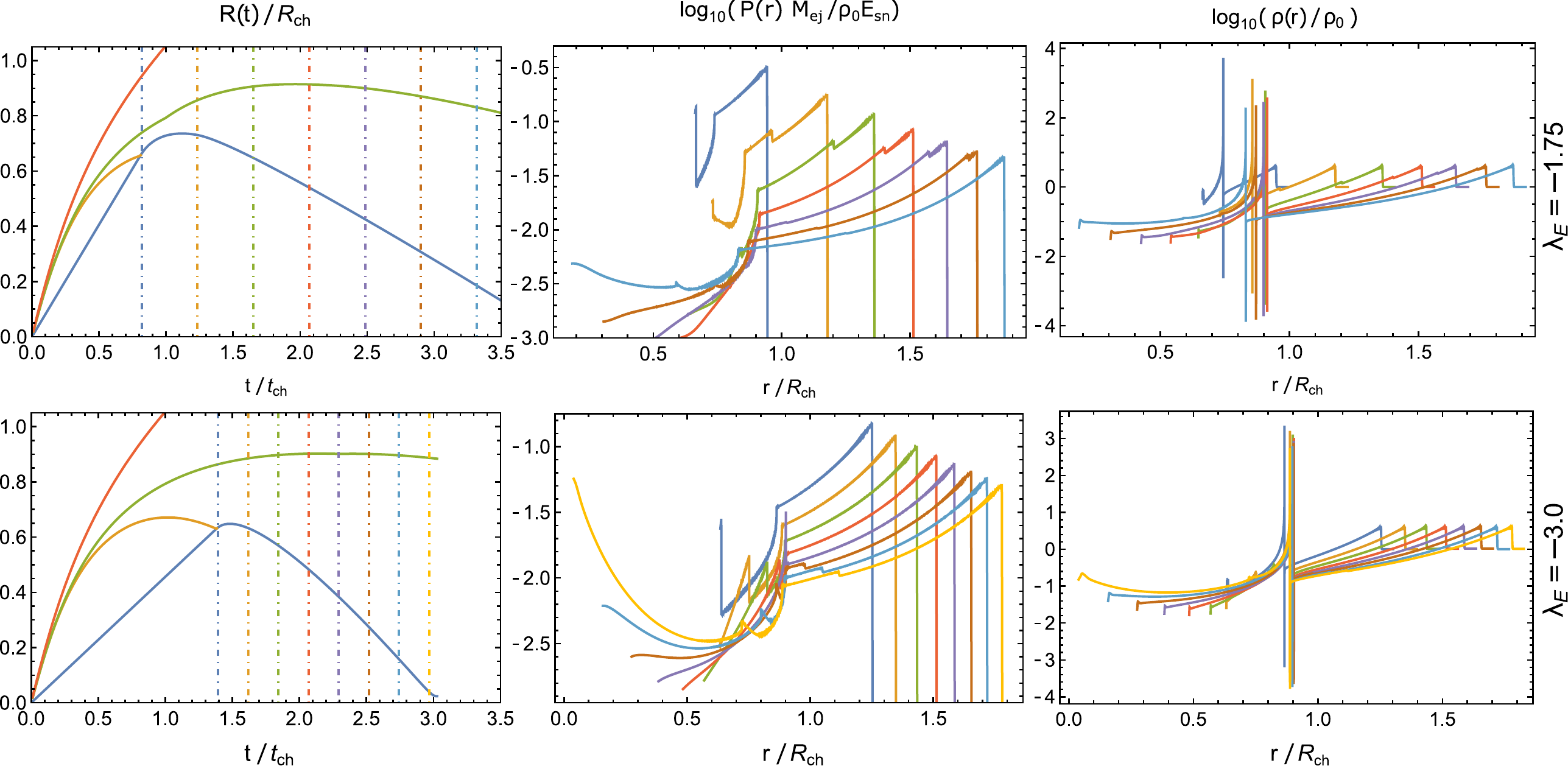}
        \caption{This figure is intended to give a global view of the time evolution of the pressure and density profiles for two specific shell+SNR models ($\lgEeff=-3.0$ and $-1.75$).
        For each model, the leftmost panel shows the radial evolution of some characteristic curves: the shell, in blue; the RS, in orange; the CD, in green; and the FS, in red. The mid panels give the pressure profiles; the rightmost panels the density ones; while the temperature profiles can be derived as the ratio of the former two. All these quantities are scaled with their respective characteristic units. The times at which these profiles are drawn are represented in the leftmost panels as dot-dashed vertical lines, with the same colour code as that used in the panels at their respective sides. 
        }
\label{fig:comparePrhoProfiles}	
\end{figure*}
The pressure (as well as density) structure inside the SNR, namely that in between the PWN shell and the FS, can be quite complex as shown in Figure~\ref{fig:comparePrhoProfiles}, where we present profiles for some selected models of shell+SNR systems since the onset of reverberation.
It is hard to capture such complexity with simplified recipes. To clarify this point, let us consider here how different choices for the outer pressure ($\Pouter$), impact the evolution of of the shell+SNR and PWN+SNR dynamics.

\begin{figure}
\centering
	\includegraphics[width=0.45\textwidth]{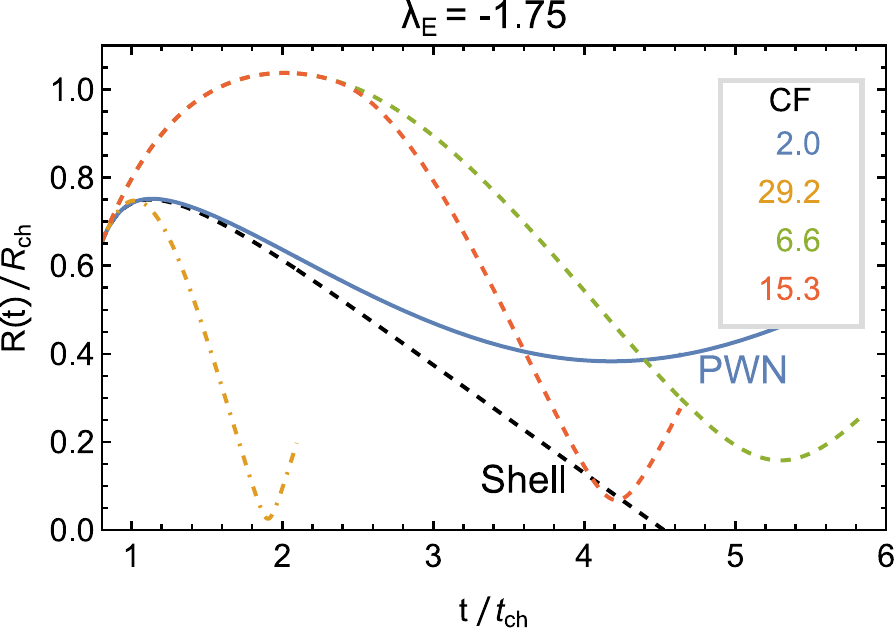}
        \caption{In this figure we show the evolution of the PWN (solid line, in blue) or shell (dotted line, in black) up to the first compression during reverberation, for the model $\lgEeff=-1.75$. For the PWN we have used $\log_{10}(\tauz/\tch)=-1.0$, while the outer pressure is given by the formulas in Appendix \ref{sec:acc-formula}. The dot-dashed line corresponds to the thin-shell model with PWN+SNR computed considering for the outer pressure $\Psedov$ from Eq.~\ref{eq:psedov}. Dashed lines indeed refer to the case with the outer pressure extrapolated from that at the RS (see Eq.~\ref{eq:pressRS}).
        In the box on the right side of the image we report the CF as computed from the different models.}
\label{fig:showCFcomp1}	
\end{figure}

\begin{figure}
\centering
	\includegraphics[width=0.45\textwidth]{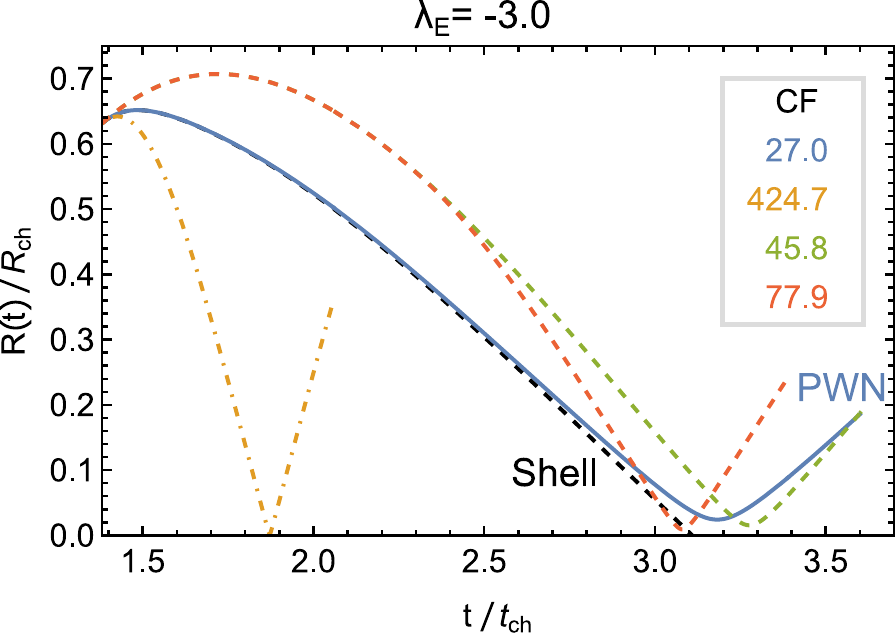}
        \caption{Same as in Fig.~\ref{fig:showCFcomp1}, but for the model 
    $\lgEeff=-3.0$. Again we have used for the PWN $\log_{10}(\tauz/\tch)=-1.0$. Colors and line stiles have the same meanings as in the previous figure.
        }
\label{fig:showCFcomp2}	
\end{figure}

Here we show a comparison between three different approaches to approximate the outer pressure to the shell: 
\begin{enumerate}[label={\arabic*.}]
    \item An outer pressure estimated using results from the Sedov model: a) either taking exactly the Sedov solution for the central pressure (as used to compute the yellow curves, in Figs.~\ref{fig:showCFcomp1}-\ref{fig:showCFcomp2}); b) or using the expansion law for the FS radius to compute the pressure downstream of the FS, and the Sedov profile to guess its value at the position of the shell (case not plotted, but similar to the previous one).
     \item The pressure immediately beyond the RS (see Eq.~\ref{eq:pressRS}, dashed lines), assuming that no PWN is present (Paper 0). This option clearly underestimates the outer pressure in the first part of the reverberation phase, because it neglects the extra pressure due to the presence of a reflected shock. On the other hand, this recipe must be arbitrarily modified before the time ($2.411\,\tch$) at which the RS reaches the center of the SNR. In order to avoid an un-physical pressure divergence, we opted to define a maximum time of validity and then to maintain the pressure constant in the following evolution. Figs.~\ref{fig:showCFcomp1} and \ref{fig:showCFcomp2} show the effect of using a maximum time of validity equal to $t_{\mathrm{implo}} - 0.3\tch$ (green curves) and $\timplo - 0.03\tch$ (red curves).
    \item The pressure profile obtained in this work using Eqs.~\ref{eq:aouter_first}-\ref{eq:Rshell_fit} and the method described in the following Appendix \ref{sec:acc-formula} (solid line);
\end{enumerate}
If we compare (see Figs.~\ref{fig:showCFcomp1} and \ref{fig:showCFcomp2}) the shell evolution during the reverberation phase, following the different prescriptions outlined above, it becomes clear that both choice 1 and 2 grossly miscalculate the evolution, even with respect to the already simplified 
shell+SNR model (black curve), which in the lower energy mode is almost coincident with the PWN+SNR model up to slightly before the first compression phase, and that in the higher energy case still provides a better estimate for the early reverberation phase. This is even more evident if one compares the CFs.

%

\section{Very early PWN evolution}
\label{sec:Vearly-pwn-evoGEN}
In the main text we have introduced a new treatment for modeling the PWN evolution under the thin-shell approximation, in the case of a flat density profile for the unshocked ejecta.
In this Appendix we present analytical results for very early times ($t\ll\tauz$) in the more general case of power-law density profiles of the unshocked ejecta, both in the core (with index $\dl$) and in the envelope (with index $\om$), namely:
\begin{equation}
\rhoej(r,t)=
\begin{cases}
A\,(\vt/r)^\dl/t^{3-\dl}, & \text{if } r < \vt t\,,	\\
A\,(\vt/r)^\om t^{\om-3}, & \text{if } r \geq \vt t\,.
\end{cases}
\label{eq:gendensityprofile}
\end{equation}
In this more general case, the parameters $A$ and $\vt$ are linked to the supernova energy $\Esn$ and to the total mass of the ejecta $\Mej$ by the relations (see Paper 0):
\begin{eqnarray}
\label{eq:vt_Agendef}
\vt = \sqrt{\frac{2\,(5-\dl)\,(\om-5)}{(3-\dl)\,(\om-3)}\frac{\Esn}{\Mej}}\,, \quad
  A = \frac{(5-\dl)\,(\om-5)}{2\upi\,(\om-\dl)}\frac{\Esn}{\vt^5}\,.
\end{eqnarray}
The case investigated in the main text corresponds to $\dl=0$ and $\om=\infty$, and for this special case formulas~\ref{eq:vt_Agendef} simplify to:
\begin{equation}
\label{eq:par_A}
    \vt=\sqrt{\frac{10}{3}\frac{\Esn}{\Mej}}\,;\qquad
    A=\frac{3}{4\upi}\frac{\Mej}{\vt^3}=\frac{5}{2\upi}\frac{\Esn}{\vt^5}\,,
\end{equation}
where $\vt$ is now the supernova boundary velocity, and the density of the ejecta is described by $\rhoej=A/t^3$.

In the general case, at times so early that it is safe to assume a constant pulsar output $\Lz$, Eqs.~\ref{eq:magfluxcons} and \ref{eq:momentumcons} can be solved analytically giving:
\begin{eqnarray}
    R_{(0)}(t)&=&\left[\frac{(3-\dl)\,(5-\dl)^3}{(9-2\dl)\,(11-2\dl)}\frac{\Lz\,t^{6-\dl}}{4\upi\,A\,\vt^\dl}\right]^{1/(5-\dl)},\quad\\
    Q_{(0)}(t)&=&\frac{5-\dl}{11-2\dl}\,\Lz\,R_{(0)}(t)\,t\,,\\
    P_{(0)}(t)&=&\frac{5-\dl}{4\upi\,(11-2\dl)}\frac{\Lz\,t}{R_{(0)}(t)^3}\,.
\end{eqnarray}
Note that this $0^{\mathrm{th}}$-order solution is equivalent to that found in \citet[][Eq. 10]{Bucciantini:2004b}.
The case of homogeneous ejecta can be also rewritten in a more convenient formulation using the two expressions for $A$ given in Eq.~\ref{eq:par_A}, namely:
\begin{equation}
R_{(0)}(t)
\simeq0.8235\,\left(\frac{\Lz\,t}{\Esn}\right)^{1/5}\!\!\!\!\vt\,t
\simeq1.5036\,\Esn^{3/10}\Mej^{-1/2}\Lz^{1/5}t^{6/5}.
\end{equation}
Also note that, in the former relation, the coefficient is slightly smaller than that (0.839) given by \citet{van-der-Swaluw:2001}.
This because here we assume that all the fluid into the shell participates of the shell velocity, while in \citet{van-der-Swaluw:2001} the authors assume that the fluid maintains the velocity reached
at the shock, finding then an average fluid velocity of  $1.15\,R(t)/t$, instead of the $1.2\,R(t)/t$ of our case.
A self-similar hydrodynamic calculation, following \citet{Jun1998}, gives an average flow velocity in the shell equal to $1.195\,R(t)/t$, where $R(t)$ is the PWN boundary, a value much closer to the one we found. Then our approximation allows for a more accurate estimate of $R(t)$, and therefore we shall adopt it in this work.

Let us now account for the time evolution of $L(t)$, by applying a power series expansion in $t$, and then by truncating it to the first order.
The evolution of $L(t)$, given in the general case by Eq.~\ref{eq:Edot}, is now: $L(t)\simeq \Lz (1-\al t/\tauz)$, and in this case the $1^{\mathrm{st}}$-order approximation of the early PWN evolution is described by the formulae:
\begin{eqnarray}
\label{eq:expfirstR}
R_{(1)}(t)\!\!\!\!&=&\!\!\!\!R_{(0)}(t)\left(1-\frac{(11-2\dl)}{(5-\dl)(49-9\dl)}\al\frac{ t}{\tauz}\right)\,, \\
\label{eq:expfirstQ}
Q_{(1)}(t)\!\!\!\!&=&\!\!\!\!Q_{(0)}(t)\left(1-\frac{(11-2\dl)(16-3\dl)}{(5-\dl)(49-9\dl)}\al\frac{t}{\tauz}\right)\,,\\
P_{(1)}(t)\!\!\!\!&=&\!\!\!\!P_{(0)}(t)\left(1-\frac{3(44-19\dl+2\dl^2)}{(5-\dl)(69-9\dl)}\al \frac{t}{\tauz}\right)\,.
\end{eqnarray}
In the lagrangian models, for mere numerical reasons, we have adopted slightly different initial conditions,
by taking a hole with size $R(\tini)$ in the ejecta distribution, and approximating the PWN at that time by assuming a linear expansion, namely:
\begin{equation}
    E\rs{PWN\,}(\tini)=\frac{\Lz\tauz^2\,(1+\tini/\tauz)^{1-\al}}{(\al-1)(\al-2)\,\tini}\left[1-\left(1+(\al-1)\frac{\tini}{\tauz}\right)\right].
\end{equation}

As better explained in the main text, with these initial conditions the very early evolution of the PWN is not strictly correct, but it converges rather rapidly to the true one.
Anyway, in order to allow an accurate comparison between lagrangian and thin-shell models, in Sec.~\ref{sec:rev-thin-shell} we have decided to use consistently these initial conditions also for the latter ones.


\section{Formulae for the outer pressure, maximum radius, and implosion velocity}
\label{sec:acc-formula}
In Sec.~\ref{sec:shell} we presented the strategy we have used to extract from our numerical outputs a formula that reasonably approximates the dynamical push from the outer SNR on a massive shell, namely $P\rs{outer}$.
Before discussing the formulae, let us remind here a few facts: since $\Pouter$ diverges when the shell implodes, we have decided to use instead the quantity $4\upi R^2\Pouter$, namely the outer force; or even better the acceleration $\aouter$, by dividing this force by the shell mass, which keeps constant during this phase.
Our goal in the fits has not been to model in detail the evolution of the pressure, i.e.\ of the acceleration, which especially in the case of higher mass (and conversely higher kinetic energy) stored in the shell show very complex behaviours, also affected by the arrival of one or more reflected shocks.
In the derivation of our fit parameters ($a$, $b$, $c$, and $k$, as shown in Eq.~\ref{eq:aouter}) we have instead tried to approximate as well as possible the evolution of the shell radius.

The results are graphically shown in Fig.~\ref{fig:parameterswholerange}, while here below we give some formulae that approximate these 4 parameters:
\begin{eqnarray}
a(\lgEeff)\!\!\!\!&=&\!\!\!\!-1.590-2.474\,\lgEeff-0.141\,\lgEeff^2+e^{-16.138-3.462\lgEeff}\nonumber\\
&&\!\!\!\!\!\!\!\!\!\!\!\!-0.0088\,e^{-56.87(\lgEeff+2.503)^2}+1.359\,e^{-30.14 (\lgEeff+1.928)^2}\nonumber\\
&&\!\!\!\!\!\!\!\!\!\!\!\!-0.864\,e^{-128.4 (\lgEeff+1.769)^2}+0.843\,e^{-27.28 (\lgEeff+1.384)^2}\nonumber\\
&&\!\!\!\!\!\!\!\!\!\!\!\!+2.630\,e^{-29.75 (\lgEeff+1.057)^2}-1.589\,e^{-177.4(\lgEeff+0.946)^2}\nonumber\\
&&\!\!\!\!\!\!\!\!\!\!\!\!+1.764\,e^{-3.621 (\lgEeff+0.514)^2};\\
b(\lgEeff)\!\!\!\!&=&\!\!\!\!-4.191-7.363\,\lgEeff-3.204\,\lgEeff^2-0.722\,\lgEeff^3 \nonumber\\
   &&\!\!\!\!\!\!\!\!\!\!\!\!+e^{-4.462\lgEeff-19.300} -0.269\,e^{-112.8 (\lgEeff+2.234)^2}\nonumber\\
   &&\!\!\!\!\!\!\!\!\!\!\!\!+3.666\,e^{-39.04 (\lgEeff+1.963)^2} +3.655\,e^{-75.18 (\lgEeff+1.847)^2}\nonumber\\
   &&\!\!\!\!\!\!\!\!\!\!\!\!-1.787\,e^{-143.0 (\lgEeff+1.756)^2} -3.642\,e^{-2.253 (\lgEeff+1.313)^2}\nonumber\\
   &&\!\!\!\!\!\!\!\!\!\!\!\!-0.181\,e^{-127.5 (\lgEeff+1.289)^2} +2.563\,e^{-69.89 (\lgEeff+1.038)^2}\nonumber\\
   &&\!\!\!\!\!\!\!\!\!\!\!\!-2.333\,e^{-166.0 (\lgEeff+0.953)^2} +5.352\,e^{-2.160 (\lgEeff+0.937)^2}\nonumber\\
   &&\!\!\!\!\!\!\!\!\!\!\!\!-1.963\,e^{-12.09 (\lgEeff+0.769)^2};\\
c(\lgEeff)\!\!\!\!&=&\!\!\!\!-1.146+0.156\,\lgEeff-0.195\,\lgEeff^2+e^{0.798\,-0.294 \lgEeff}\nonumber\\
   &&\!\!\!\!\!\!\!\!\!\!\!\!-0.015\,e^{-410.8 (\lgEeff+2.264)^2}+0.244\,e^{-28.23 (\lgEeff+1.910)^2}\nonumber\\
   &&\!\!\!\!\!\!\!\!\!\!\!\!+0.167\,e^{-42.51 (\lgEeff+1.469)^2}+0.599\,e^{-16.63 (\lgEeff+1.045)^2}\nonumber\\
   &&\!\!\!\!\!\!\!\!\!\!\!\!-0.408\,e^{-177.3 (\lgEeff+0.936)^2}+0.206\,e^{-52.05 (\lgEeff+0.609)^2}\nonumber\\
   &&\!\!\!\!\!\!\!\!\!\!\!\!+0.592 e^{-100.8 (\lgEeff+0.377)^2};\\
k(\lgEeff)\!\!\!\!&=&\!\!\!\!0.586-0.582\,\lgEeff-0.129\,\lgEeff^2+e^{-0.302-0.592\lgEeff}\nonumber\\
   &&\!\!\!\!\!\!\!\!\!\!\!\!-0.362\,e^{-102.7 (\lgEeff+2.235)^2}+2.780\,e^{-19.101 (\lgEeff+1.803)^2}\nonumber\\
   &&\!\!\!\!\!\!\!\!\!\!\!\!-1.503\,e^{-102.4 (\lgEeff+1.683)^2}-0.867\,e^{-9.655 (\lgEeff+1.555)^2}\nonumber\\
   &&\!\!\!\!\!\!\!\!\!\!\!\!+1.149\,e^{-54.465 (\lgEeff+1.031)^2}-1.049\,e^{-191.5 (\lgEeff+0.933)^2}.
\end{eqnarray}
Only for very small energies ($\lgEeff<-4.5$) we have adopted a simpler expression for $\aouter$, by using a pure exponential, $\aouter(x)=\exp(c\rs{low}-k\rs{low} x)$, with:
\begin{eqnarray}
c\rs{low}(\lgEeff)\!\!\!\!&=&\!\!\!\!1.025+(0.08807+0.01065\lgEeff)e^{-0.755\lgEeff};\quad\\
k\rs{low}(\lgEeff)\!\!\!\!&=&\!\!\!\!3.149+(0.1680+0.0168\lgEeff)e^{-0.950\lgEeff}.
\end{eqnarray}
For our simplified thin-shell model, as from Sec. ~\ref{sec:simplified}, we need instead the interpolations of the maximum size ($\Rmax$) and of the asymptotic velocity of implosion ($\Vimplo$).
Here are analytic interpolations for these two quantities:
\begin{eqnarray}
\Rmax(\lgEeff)\!\!\!\!&=&\!\!\!\!
0.8650-0.0313\lgEeff-0.0448\lgEeff^2-0.0038\lgEeff^3\nonumber\\
&&\!\!\!\!\!\!\!\!\!\!\!\!-0.051 e^{-1771.(10.^\lgEeff-0.0398)^2}-0.050 e^{-0.866(1.434+\lgEeff)^2}\nonumber\\
\label{eq:Rmaxapprox}
&&\!\!\!\!\!\!\!\!\!\!\!\!-0.024 e^{-56.31(1.310+\lgEeff)^2}-0.019 e^{-146.8(0.883+\lgEeff)^2};\\
\Vimplo(\lgEeff)\!\!\!\!&=&\!\!\!\!
\label{eq:Vimploapprox}
-0.078-0.1883\lgEeff+0.00059\lgEeff^2+e^{-1.259+1.232\lgEeff}\nonumber\\
&&\!\!\!\!\!\!\!\!\!\!\!\!+0.0326 e^{-21.86(2.217+\lgEeff)^2}-0.0926 e^{-7.747(1.568+\lgEeff)^2)}\nonumber\\
&&\!\!\!\!\!\!\!\!\!\!\!\!
+0.0434 e^{-48.38(1.286+\lgEeff)^2}+0.080 e^{-0.3360(1.241+\lgEeff)^2}
\nonumber\\
&&\!\!\!\!\!\!\!\!\!\!\!\!-0.053 e^{-57.43(0.939+\lgEeff)^2}+0.048 e^{-124.0(0.918+\lgEeff)^2}.
\end{eqnarray}

\section{PWN evolution for different braking indices}
\label{sec:pwn-evo-long}
\begin{figure}
\centering
	\includegraphics[width=0.45\textwidth]{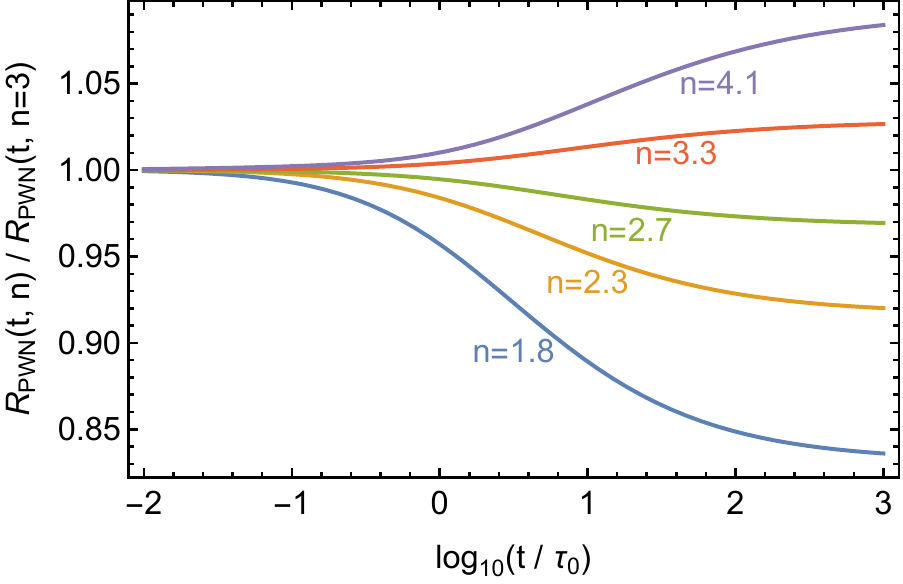}
        \caption{Comparative evolution of the PWN radius for different values of the the braking index $n$, divided by the corresponding solution for $n=3$ (magnetic dipole model).
        }
\label{fig:Comparedifferentn}	
\end{figure}

While in this paper for the PWN evolution we have considered only the case with braking index $n=2.33$, corresponding to a fading index $\al=2.5$, in this Appendix we present analytic formulae that are excellent approximations over a wider range.
For simplicity, we will consider here only the case of homogeneous ejecta ($\dl=0$).
As for the functional form of the analytic approximation for $\Rpwn(t)$, we have found that Eq.~\ref{eq:Rapprox} performs very well, and then we have used it.
This formula has 4 parameters one of which, $\CRz$, is independent of value of the braking index and can be determined analytically as $\simeq0.7868$ (see Eq.~\ref{eq:solrt_A}).
As for the other 3 parameters, we have run thin-shell numerical models with a finely spacing in $n$, and then we have performed a polynomial fit on them.
For $n$ in the range $[1.8,\,4.1]$, which includes both our case and that of the pure dipole braking, we have obtained accurate interpolations of the parameters by using third-degree polynomials:
\begin{eqnarray}
  \CRinf(n)\!\!\!\!\!\!&\simeq&\!\!\!\!\!\!0.30139+0.46268 n-0.099087 n^2+0.008715 n^3,\quad \\
    a(n)\!\!\!\!\!\!&\simeq&\!\!\!\!\!\!0.89882-0.00365 n-0.045432 n^2+0.006836 n^3,\\
      b(n)\!\!\!\!\!\!&\simeq&\!\!\!\!\!\!0.78755+0.08107 n-0.068173 n^2+0.008969 n^3.
\end{eqnarray}
With these formulae the maximum deviation from the numerical profiles for $\Rpwn(t)$ keeps always $\lesssim 1\%$ in the considered range ($\lesssim0.1\%$ if one considers $[2.3,\, 3.3]$).
Fig.~\ref{fig:Comparedifferentn} shows the time evolution of the PWN size for different values of the breaking index, scaled with the case $n=3$.
One may notice that the differences are rather minor, amounting to just a few percent.

\label{lastpage}

\end{document}